\documentclass[manuscript]{aastex}
\usepackage[export]{adjustbox}
\usepackage{lineno}

\shorttitle{Observations of the chromospheric Evershed flow of sunspot penumbra}
\shortauthors{Romano, Schillir\'{o} and Falco}

\begin{document}


\title{Observations of the chromospheric Evershed flow of sunspot penumbra by the application of the SOM technique}

\author{P. Romano\altaffilmark{1}, F. Schillir\'{o}\altaffilmark{1} and M. Falco\altaffilmark{1}}

\email{paolo.romano@inaf.it}




\altaffiltext{1}{INAF - Osservatorio Astrofisico di Catania,
              Via S. Sofia 78, 95123 Catania, Italy.}



\begin{abstract}
 The sunspot penumbra is usually observed in the photosphere and it is of particular interest for its magneto-convection which seems to transport the heat from the top of the convection zone into the solar atmosphere. It is well known that the penumbra magnetic field extends in the upper layers of the solar atmosphere forming the so called super-penumbra. Thanks to the application of the Self Organizing Map technique to a spectral dataset containing monochromatic images acquired along the Ca II 854.2 nm and H$\alpha$ 656.28 nm lines, we were able to segment the penumbra and to measure the plasma velocity along the chromospheric portions of penumbral filaments. We found that the head, body and tail of penumbral filaments show vertical flows compatible with the persistence of the Evershed flow. Instead, the inverse Evershed flow has been observed only in the outer portion of the super-penumbra. We found that two opposite Evershed regimes work next to each other, without overlapping and both contribute to the downflow around sunspots. These results confirm the uncombed model of the sunspot penumbra and provide some hints that the downflow around sunspots may be ascribed to the magnetic field dragging the plasma down.

\end{abstract}




\section{Introduction}

The region surrounding sunspot "umbra" in a fully developed sunspot is known as penumbra. This is an interesting physical system because in its fine structure the conditions of the plasma and magnetic fields are quite particular. Indeed the plasma temperature is intermediate in comparison with the quiet Sun and the umbra of the sunspots, while the magnetic field, decreasing its intensity as function of the umbra distance, shows an uncombed configuration \citep{Sol93}. These peculiar physical conditions of the penumbra need to be investigated with more detail using high resolution data and more and more sophisticated methods of analysis \citep[for a review about recent results concerning properties of the penumbrae see, e.g.,][]{Sol03, Borr11, Tiw17, Al-19}.

Recent progresses in the observations of the penumbra have been obtained by Hinode \citep{Tiw13}. The high resolution data of the Solar Optical Telescope (SOT) onboard of the satellite allowed to isolate some important constraints in the physical properties of the penumbra. The inner footpoints of the penumbral filaments showed a larger brightness in comparison with their outer edges, while the magnetic field inclination appeared like a strongly flattned $\bigcap$-loop. Moreover, the line-of-sight (LOS) velocity observed along each filament appeared as upflow and downflow in the inner and outer edges, respectively. These velocity fields have been interpreted as the result of the interaction between the penumbra magnetic field and the more vertical background magnetic fields, where the outwards acceleration of the plasma along the filament axis, i.e., Evershed flow \citep{Eve09}, is maintained by a gas pressure gradient \citep{Sch05}.

Actually, SOT/Hinode data showed also an azimuthal component of the horizontal velocity along the whole axis of those filaments. The presence of azimuthal convection throughout the penumbra is closer to the idea of \citet{Spr06}, where the penumbra is interpreted as formed by bright filaments characterized by field-free gap on a dark background \citep{Sch06}. Moreover, \citet{Tiw13} discovered that the magnetic field strength is greater at the tails of penumbral filaments compared to the heads of the filaments, on average. This implies that the magnetic pressure is higher at the heads than that on the tails. This configuration could potentially facilitate a siphon flow mechanism, which in turn could contribute to the generation of the Evershed flow. Therefore, the possibility of a siphon flow playing a role in the observed phenomena was not excluded by the authors.

In the field of the MHD simulations some important progresses have been obtained to reproduce many aspects of the fine structure of the penumbra. In particular, some of them \citep{Rem11, Rem12, Rem15} seems to be able to interpret the magnetic field of the penumbral filaments and their association with horizontal flows close to the solar surface as compatible with the embedded flux model \citep{Sol93, Sch98a, Sch98b}. This model is based on the assumption that the flux tube forming a penumbral filament is heated by the radiation coming from the underlying hotter quiet Sun, therefore, expanding and getting less dense than the surroundings, floats in the solar atmosphere \citep{Bel03, Bel04}. However, there are several evidences that the penumbra forms, not directly from the buoyancy of the flux tubes, but from their changes in the inclination in the magnetic canopy overlying the pore and returning beneath the photosphere  \citep{Rom13, Rom14, Rom20}.

Moreover, the interpretation of the plasma flows along the penumbral filaments and the corresponding magnetic field inclination in the context of the explanation of the heat transport is not trivial. Indeed, the penumbra bolometric brightness (about 75\% of the quiet photosphere on average) should require at least vertical velocities of about 1-2 km s$^{-1}$ to justify the heat transport from the sub-photospheric zone, but these velocities are not observed \citep{Lan05}.

Threfore, further investigations of the physical properties of the sunspot penumbrae are necessary to reconstruct the behaviour of the plasma in their peculiar configuration of the magnetic field. However, it is important to also investigate the penumbra properties higher in the chromospheric layer where radially elongated fibrils, visually resemble an enlarged version of the photospheric penumbra. These fibrils around a sunspot are named chromospheric “super-penumbra” \citep{Lou68}. They are presumably aligned with magnetic field lines and connecting much further into the solar surface than a typical sunspot penumbra \citep{Jin19}. Although, high resolution observations are necessary to deep the physical properties of the coupling between magnetic field and plasma in those structures, it is well known that the chromospheric super-penumbra hosts the inverse Evershed flow \citep{Mor95}, i.e., a nearly horizontal mass inflow toward the umbra, combined with a downflow near the umbral border. Further observations of the plasma flow using different chromospheric line are useful to provide significant constraints to model the whole sunspot penumbra at different altitudes and stages.

To reach this goal new segmentation methods are useful to isolate the different portions of the thin and radially elongated filaments. In this paper, we describe the results obtained by the analysis of high resolution spectroscopic data applying the Self Organizing Map technique \citep[SOM;][]{Sch21} as a tool for the segmentation of the chromospheric penumbra fine structure. Thanks to this machine learning approach we are able to deduce the properties of the chromospheric portion of the sunspot penumbra, providing new results which are able to complete the scenario of the coupling between plasma e magnetic field in this portion of the solar sunspots. After the description of the data used for this analysis reported in the Section 2, we highlight the advantage of the SOM technique in this kind of studies in Section 3. Then, in Section 4 we describe the obtained results, while we discuss our conclusions in Section 5.


\section{Data}
\begin{figure}
\begin{center}
\includegraphics[trim=0 140 160 75, clip, scale=0.35]{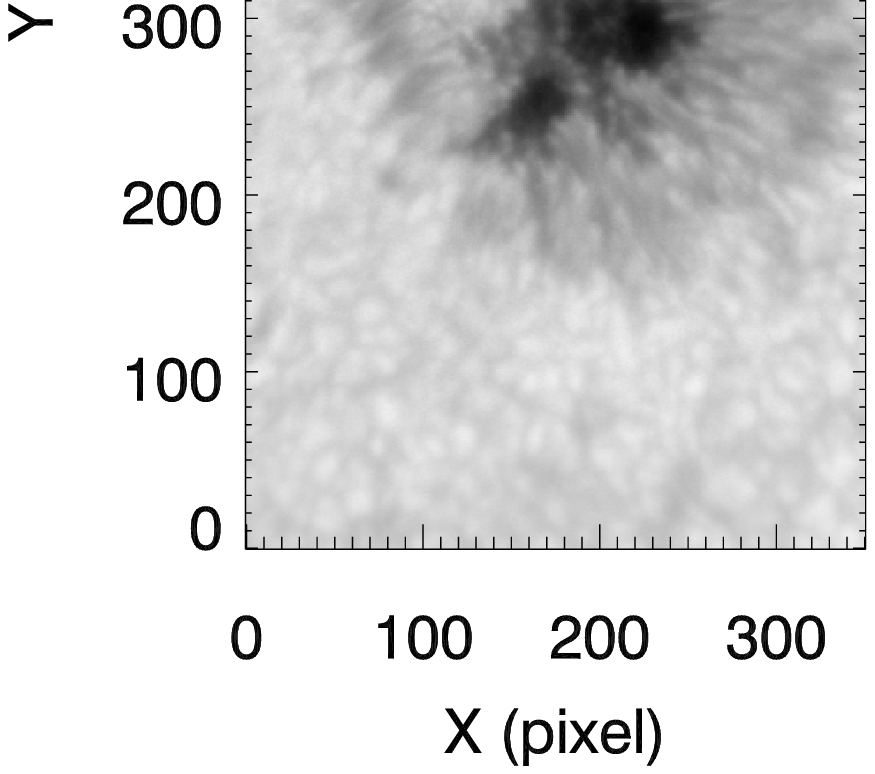}
\includegraphics[trim=110 140 160 75, clip, scale=0.35]{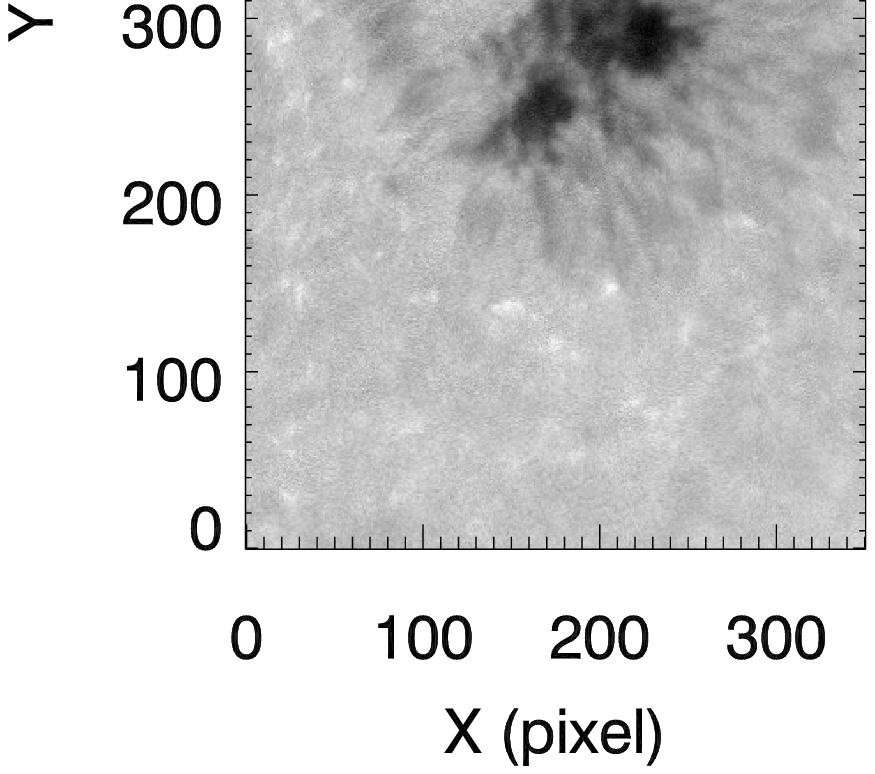}
\includegraphics[trim=110 140 160 75, clip, scale=0.35]{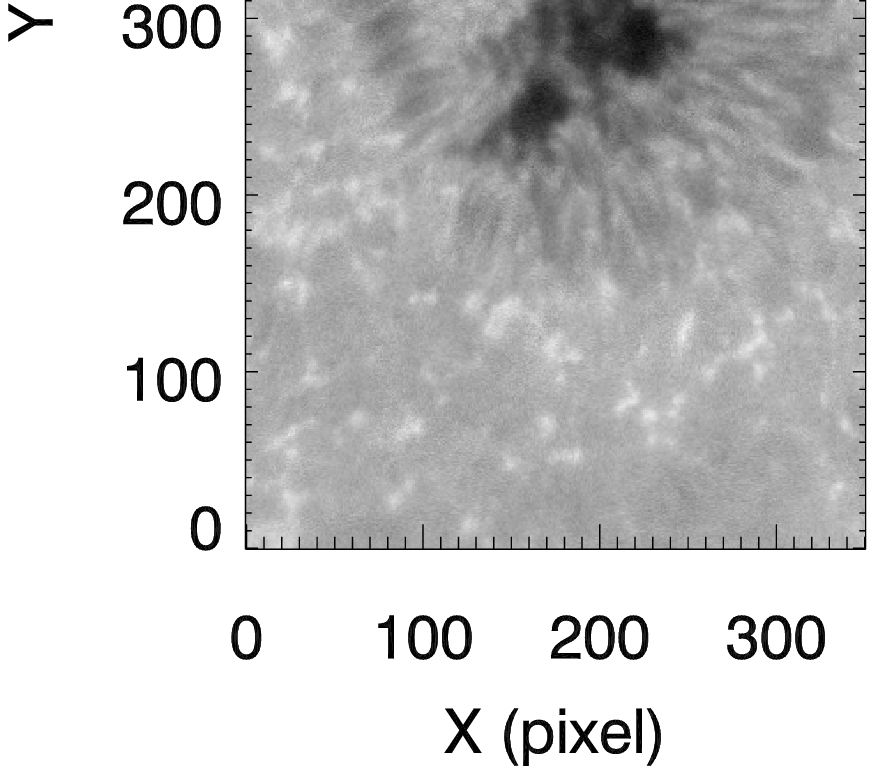}\\
\includegraphics[trim=0 60 160 75, clip, scale=0.35]{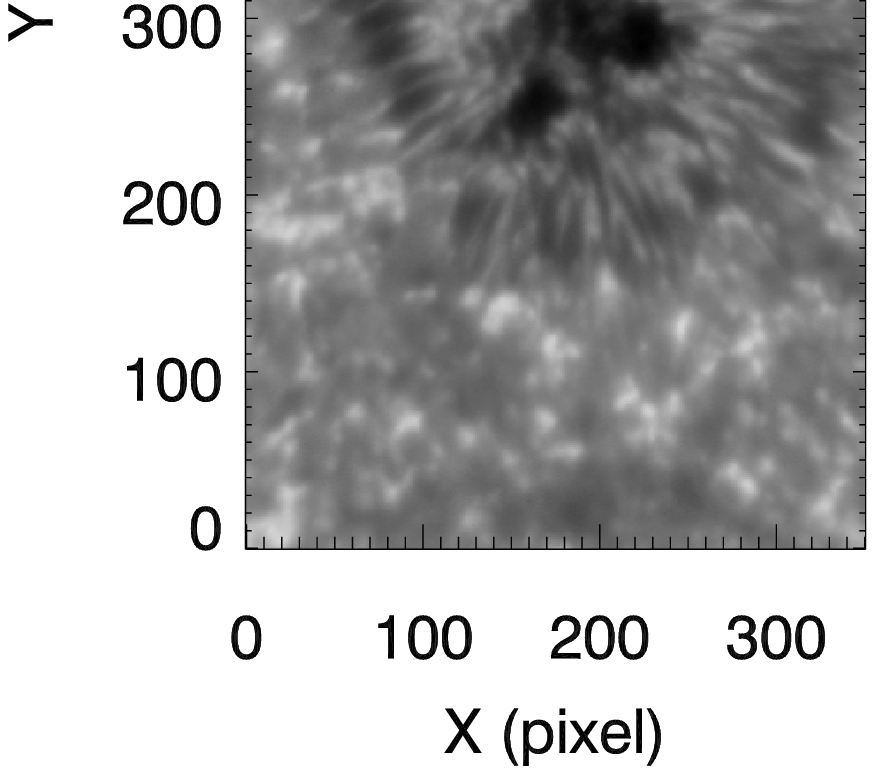}
\includegraphics[trim=110 60 160 75, clip, scale=0.35]{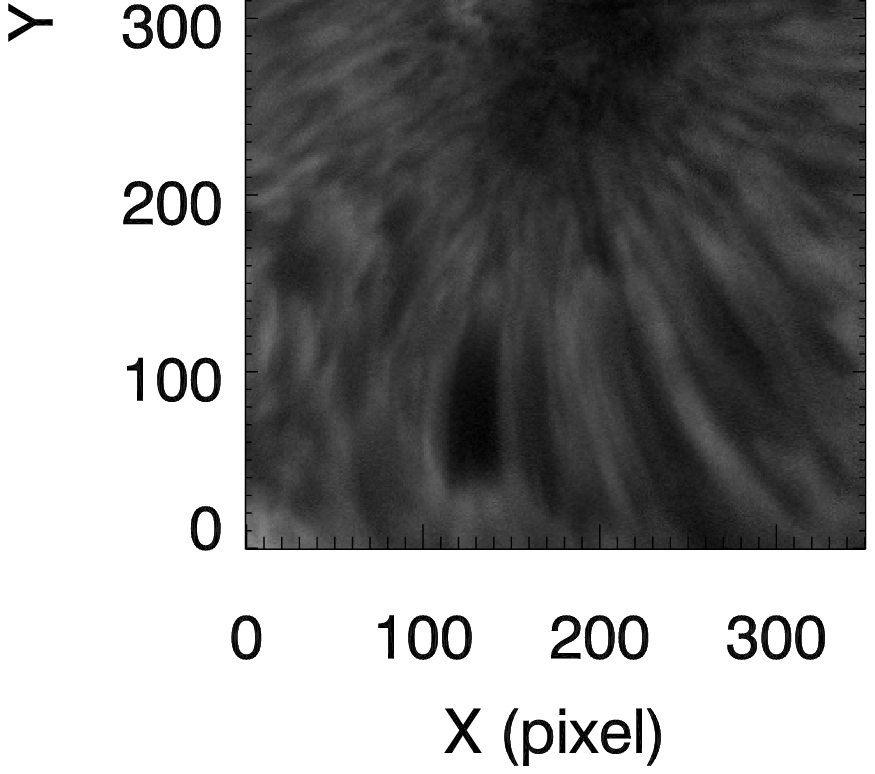}
\includegraphics[trim=110 60 160 75, clip, scale=0.35]{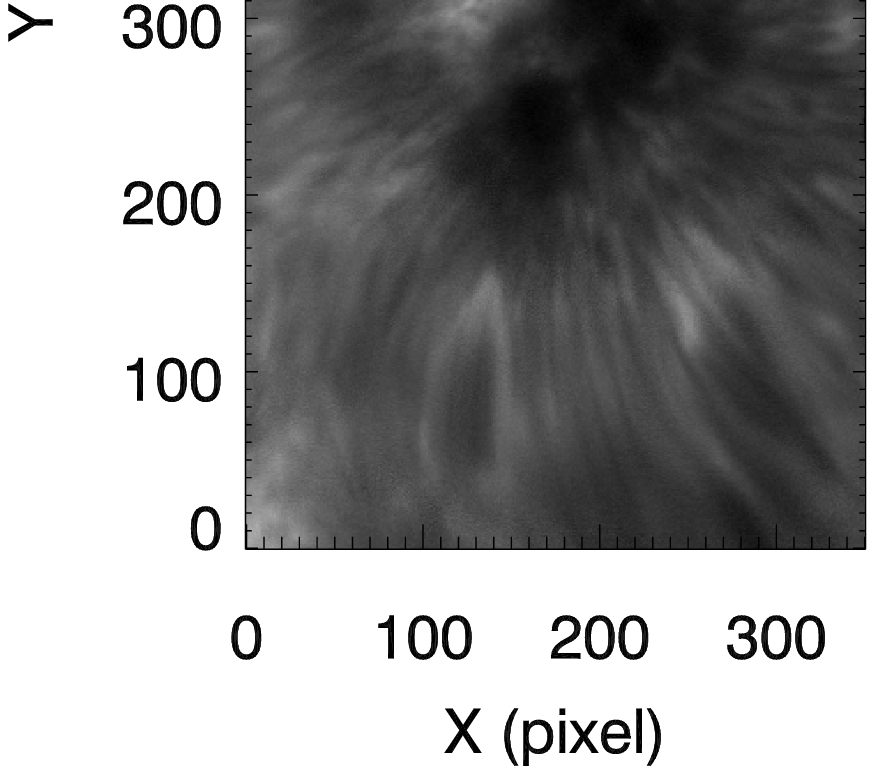}\\
\caption{Spectral images acquired by IBIS in the blue wings (top panels) and in the core (bottom panels) of the Fe I 630.250 nm (left panels), H$\alpha$ (middle panels), and Ca II (right panels) lines. The white box in the bottom left panel indicates the field of view reported in Figure xx, where an example of dark penumbral filament is shown. All the images have been acquired on May 18 at 14:42 UT.}
\label{Fig0}
\end{center}
\end{figure}

The high-resolution data, with a pixel scale of 0\farcs095, taken by the Interferometric Bidimensional Spectroscopic Instrument \citep[IBIS;][]{Cav06} on 2015, May 18, were used to investigate in detail the morphology at the chromospheric level of the panumbra surrounding the main sunspot of the AR NOAA 12348, which was located next to the central meridian (S10 E10). At that time IBIS was operating at the NSO/Dunn Solar Telescope (DST) and the Catania Solar Telescope \citep{Rom22} was used as auxiliary instrument to find the targets of the observing campaign. Monochromatic images along one photospheric line (Fe I 630.250 nm) and two chromospheric lines (Ca II 854.2 nm and H$\alpha$ 656.28 nm lines) have been acquired during that observing campaign. The cadence of each scan was about 67 s. All of these lines were sampled with a spectral FWHM of 2 pm, an average spectral step of 2 pm, and an integration time of 60 ms. The datasets of the photospheric line contained narrow band images in 30 spectral points (see left panels of Fig.\ref{Fig0}), while the dataset of the two chromospherc lines contained narrow band images in 17 spectral points along the H$\alpha$ 656.28 nm line (see middle panels of Fig.\ref{Fig0}) and in 25 spectral points along the Ca II 854.2 nm line (see right panels of Fig.\ref{Fig0}) \citep[see][for further details]{Rom17}. The field of view (FOV) of IBIS was 500$\times$1000 pixels, but we considered often in our analysis only a sub-FOV of 350$\times$700 pixels centered on the preceding sunspot of the AR. We also acquired broadband images at 633.3 nm simultaneously with the spectral frames, imaging the same FOV with the same exposure time. The spectra have been normalized to the quiet Sun continuum, Ic. We restored images using the Multi-frame Blind Deconvolution \citep[MFBD;][]{Lof02} technique to reduce the seeing degradation, achiving a spatial resolution of about 0\farcs25 at 630.250 nm. Such kind of data, stored in the IBIS data Archive \citep[IBIS-A;][]{Erm22}, proved to be very useful for the investigation of the penumbra properties thanks to their high spatial, spectral and temporal resolution \citep[see][]{Rom13, Rom14, Mur16, Mur17, Rom17, Rom20}. 

As we can see in Fig. \ref{Fig0}, the super-penumbra fibrils {in the core of the Ca II  and H$\alpha$ lines} along the East-West direction are not fully contained into the IBIS FOV, although the filamentary pattern is clear and visible especially along the North-South direction. The comparison between the two chromospheric lines highlights some morphological differences of the penumbra filaments and super-penumbra fibrils. The penumbra fine structure exhibits remarkable similarities in terms of shape and brightness variations among the three spectral lines. For instance, as illustrated in the top panels of Fig. \ref{Fig0}, the arrows indicate a specific penumbral filament that is visible in the blue wings of all three spectral lines. This consistent appearance of the penumbral filament across the different spectral lines suggests a close correlation in the underlying physical processes and atmospheric conditions within the penumbra, which are being probed by the respective spectral lines. Instead, the H$\alpha$ super-penumbra fibrils  (bottom middle panel of Fig. \ref{Fig0} seem more abundant and contrasted than the ones observed in the center of the Ca II line (bottom right panel of Fig. \ref{Fig0}). It is noteworthy that in the center of the Ca II line some inner footpoints of the wider super-penumbra fibrils show some brightenings. Moreover, as the wavelength moves to the wings of the lines, the super-penumbral fibrils become more and more discrete and sparse. The correspondence of some super-penumbral fibrils in the two spectral lines can be recognized, especially the curled filament that is located in the northern part of the FOV and intrudes the sunspot next to its light bridge \citep{Gug19}.


\section{Analysis}

In order to analyse the vertical velocity in different portions of the penumbra a segmentation of the images taken at different wavelength was necessary. In particular, to study the variation of the physical properties along the radial direction of penumbral filaments we had to distinguish between their head, body and tail \citep[see][]{Tiw15}. This task is very complicate due to the strong variations of the penumbra morphology and brightness as the wavelength varies in our spectral dataset. For this reason we applied for one of the first times the SOM technique to our high resolution photospheric and chromospheric images taken by IBIS. In this way we found that the SOM technique allowed to exploit the plenty of information contained in spectral images acquired by the new generation solar instruments dedicated to the acquisition of several narrow band images along photospheric and the chromospheric spectral lines with a fast rate.   

The self-organization algorithm used in our study consists of several main processes. It begins with an input layer, which serves as the source of the data to be processed. The algorithm also includes a fully connected neural lattice, which adapts its structure and morphology to replicate the organization of the input data based on their neighborhood relationships. In our case, this lattice morphology is employed to group pixels into regions or features. These features are detected based on the distance between the multispectral data collected for each pixel and the normalized random initial formatting of the neural lattice. 
More details about the SOM technique and its application to high resolution data taken by a bidimensional spectrometer are described in detail in \citet{Sch21}. We remark that the SOM is an important class of neural networks capable of unsupervised autonomous learning from data, inspired by neuro-biological studies. In our case, we applied a particular kind of SOM: the Kohonen Network \citep{Koh01}, belonging to the class of vector-coding algorithms and allowing data compression, data clustering or data classification.

\begin{figure}
\begin{center}
\includegraphics[trim=0 20 160 75, clip, scale=0.35]{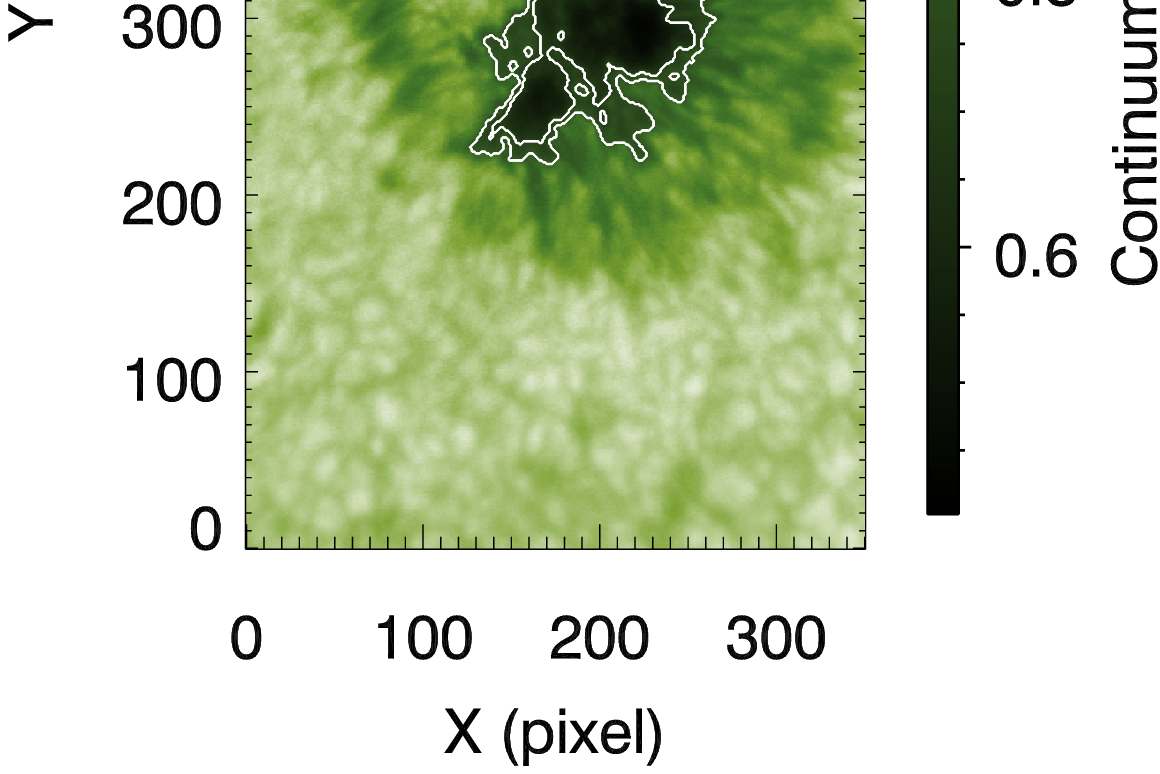}
\includegraphics[trim=110 20 160 75, clip, scale=0.35]{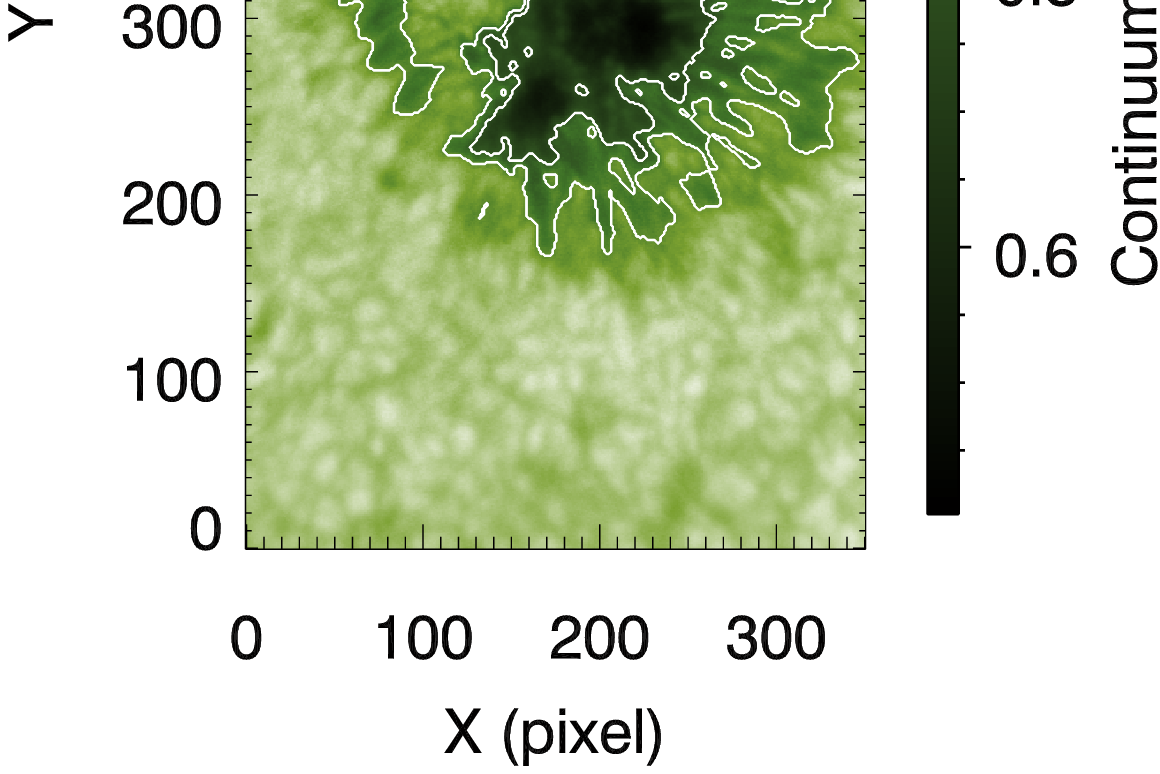}
\includegraphics[trim=110 20 0 75, clip, scale=0.35]{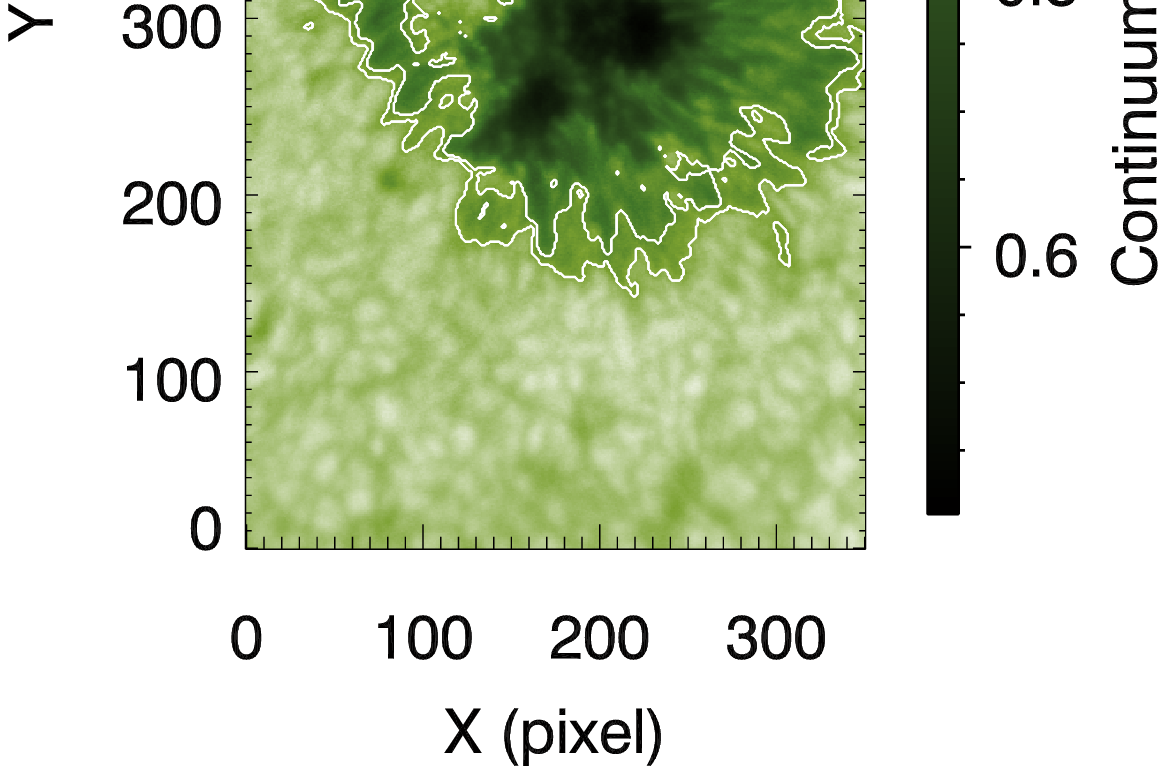}
\caption{Contours indicating the heads (left panel), the bodies (central panel) and the tails (right panel) of the penumbral filaments as identified by the SOM technique applied to the Fe I line at 630.250 nm. The contours of the features are overplotted to the images taken in the first spectral point of the line scan, i.e. 630.218 nm.}
\label{Fig_test}
\end{center}
\end{figure}

\begin{figure}
\begin{center}
\includegraphics[trim=0 20 160 75, clip, scale=0.35]{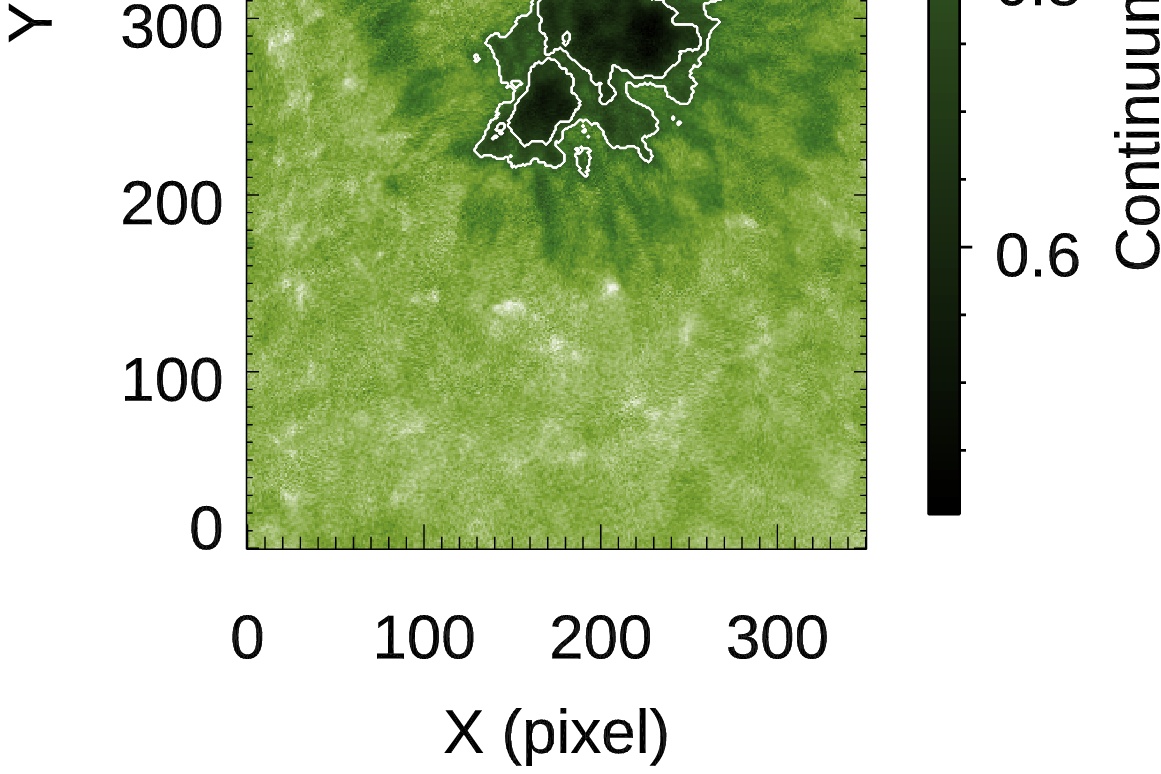}
\includegraphics[trim=110 20 160 75, clip, scale=0.35]{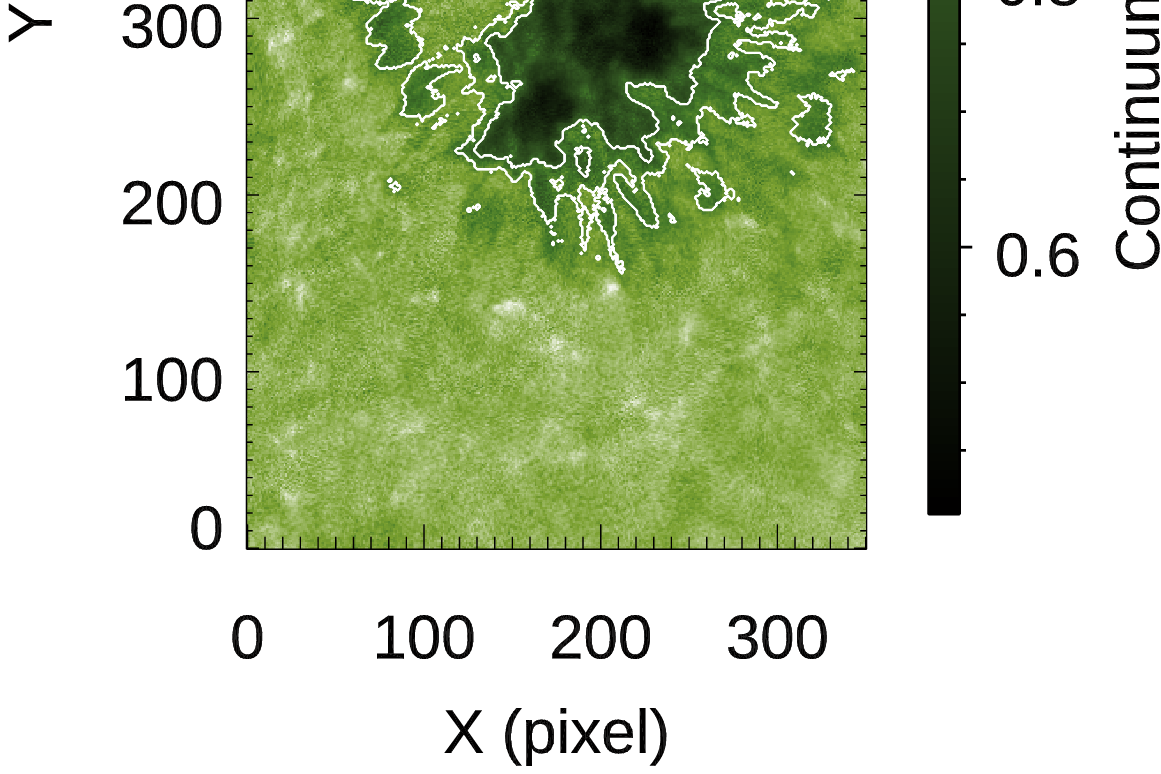}
\includegraphics[trim=110 20 0 75, clip, scale=0.35]{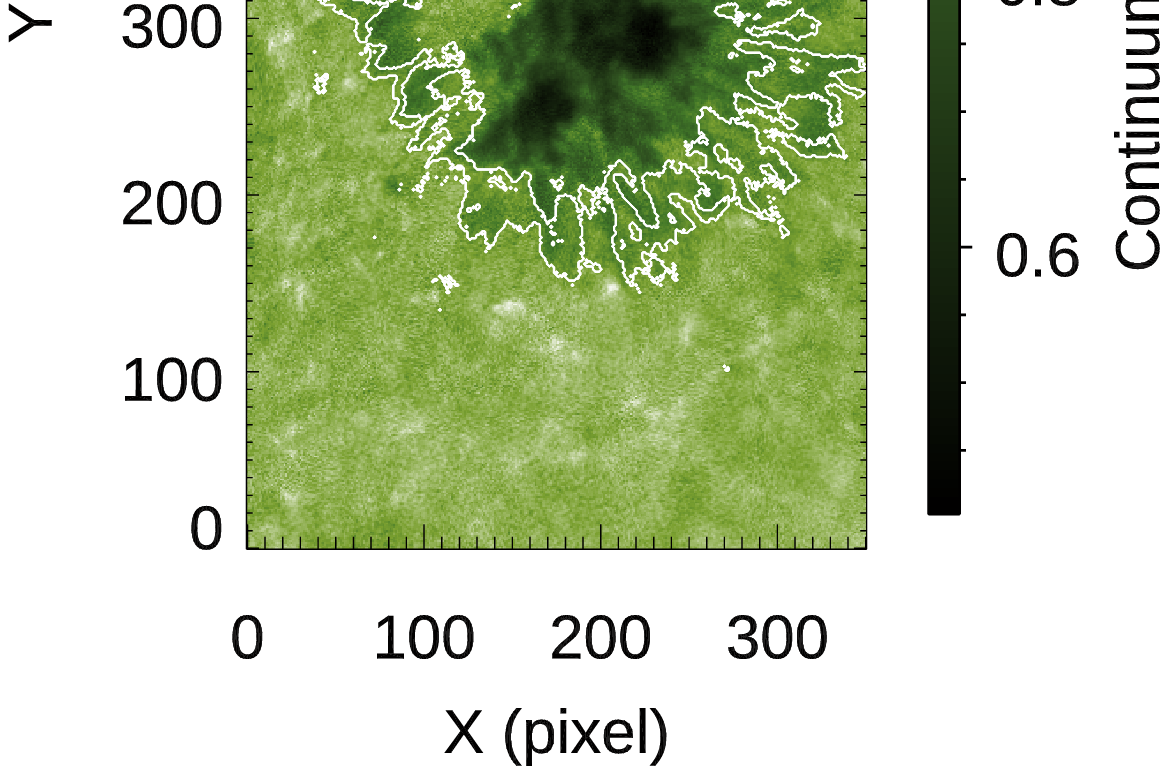}
\caption{Contours indicating the heads (left panel), the bodies (central panel) and the tails (right panel) of the penumbral filaments as identified by the SOM technique applied to the H$\alpha$ line. The contours of the features are overplotted to the images taken in the first spectral point of the line scan, i.e. 656.134 nm.}
\label{Fig1}
\end{center}
\end{figure}

\begin{figure}
\begin{center}
\includegraphics[trim=0 20 160 75, clip, scale=0.35]{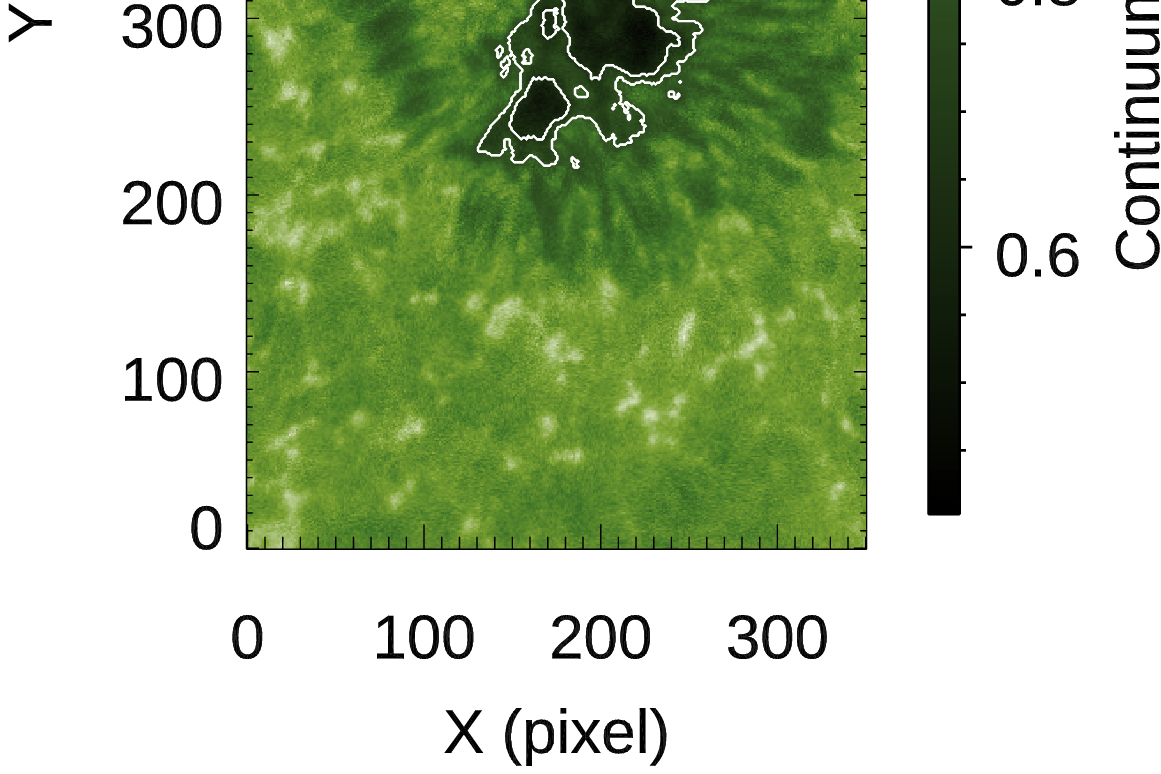}
\includegraphics[trim=110 20 160 75, clip, scale=0.35]{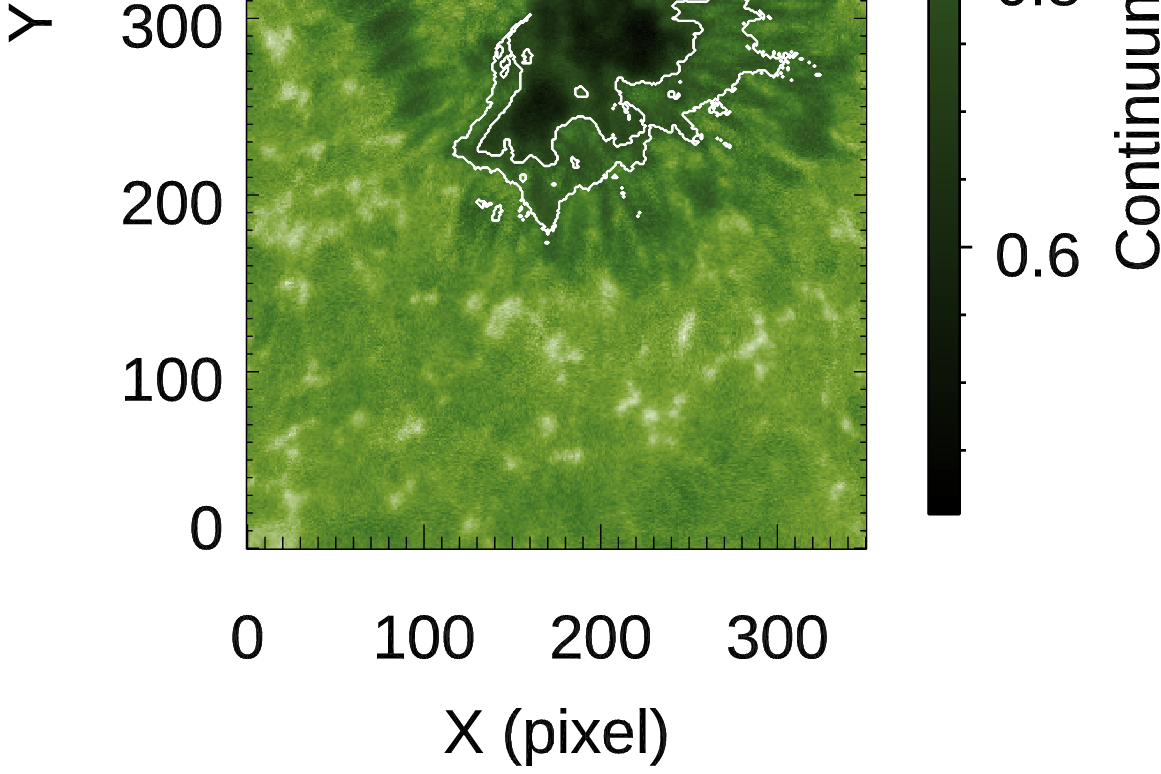}
\includegraphics[trim=110 20 0 75, clip, scale=0.35]{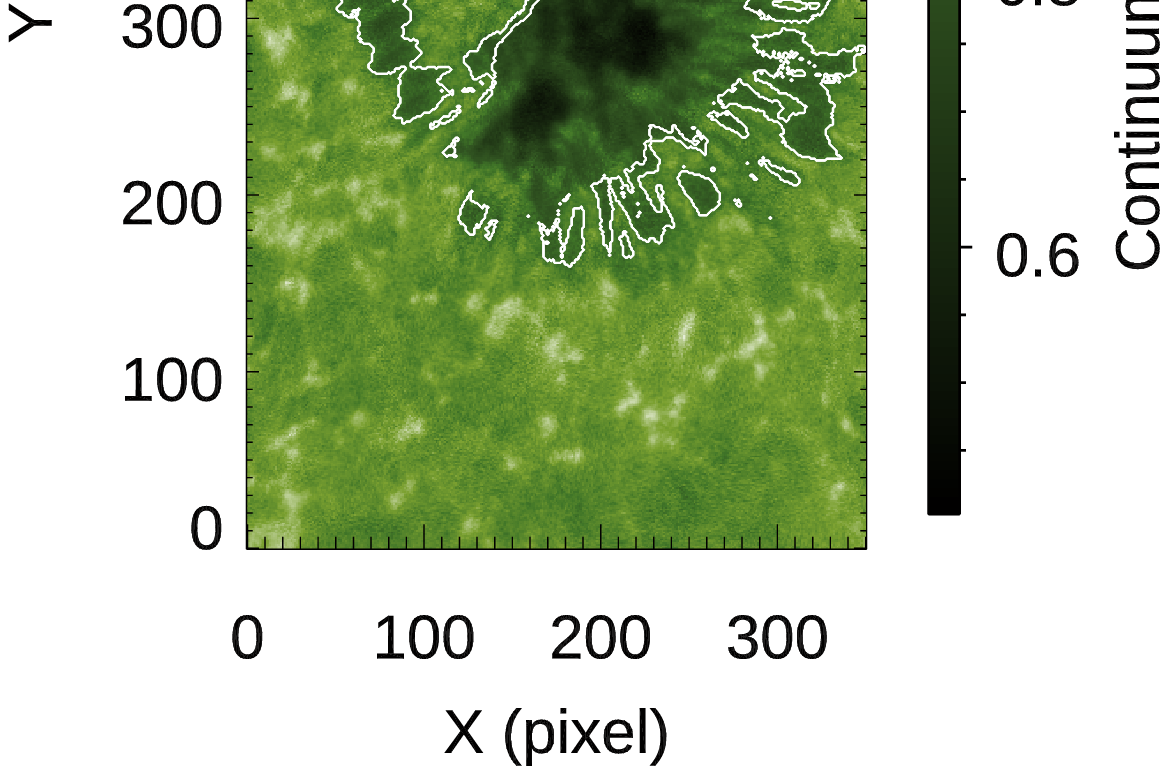}
\caption{Contours indicating the heads (left panel), the bodies (central panel) and the tails (right panel) of the penumbral filaments as identified by the SOM technique applied to the Ca II line. The contours of the features are overplotted to the images taken in the first spectral point of the line scan, i.e. 854.071 nm.}
\label{Fig2}
\end{center}
\end{figure}

Therefore, the application of the SOM allowed the image segmentation by the association of each pixel of each frame to a unique cluster, named feature. It is noteworthy that the SOM network is organized to guarantee segmentation in complementary regions, so that the results of the elaboration are two-dimensional maps, which display the features, i.e., semantically separated regions. Using an unsupervised approach the number of features was not previously specified, and it came from data structure as a results of the data processing. According to \citet{Sch21} we decided to consider the results obtained for 16 clusters for each spectral line. For the scope of this work we used only nine features (three per spectral line) corresponding to the heads, the bodies and the tails of the penumbral filaments in the Fe I 630.25 nm (Figure \ref{Fig_test}), H$\alpha$ 656.28 nm (Figure \ref{Fig1}) and Ca II 854.2 nm (Figure \ref{Fig2}) lines. We found some slight differences between the clastering of the three datasets, although in all cases we obtained a very good segmentation of the penumbra, as we can see by the contours of the features overplotted to the images taken in the first spectral points of each line scan, i.e. corresponding to the blue wings of the lines where the photospheric structures are more visible. The primary advantage of this technique is that it enables the segmentation of specific fine structures of interest without requiring user supervision, while disregarding other structures that differ, such as the curled filament that intersects the penumbra near the light bridge.

\begin{figure}
\begin{center}
\includegraphics[trim=0 20 160 75, clip, scale=0.35]{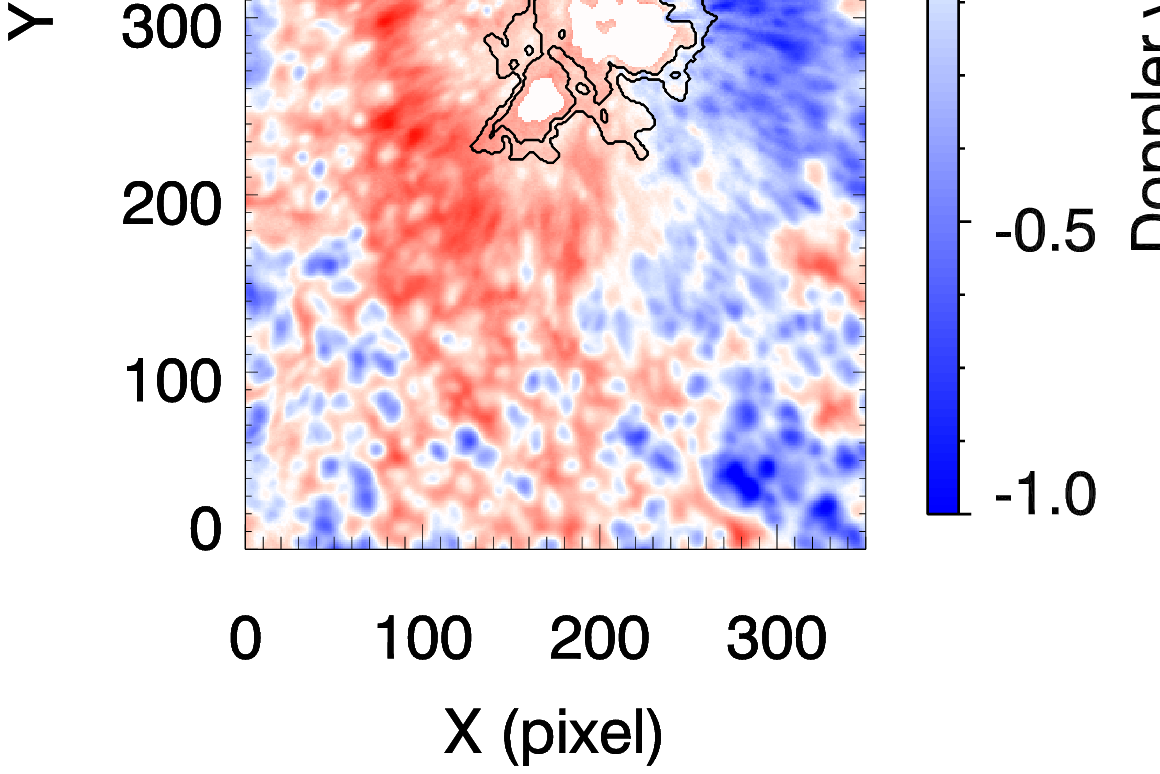}
\includegraphics[trim=110 20 160 75, clip, scale=0.35]{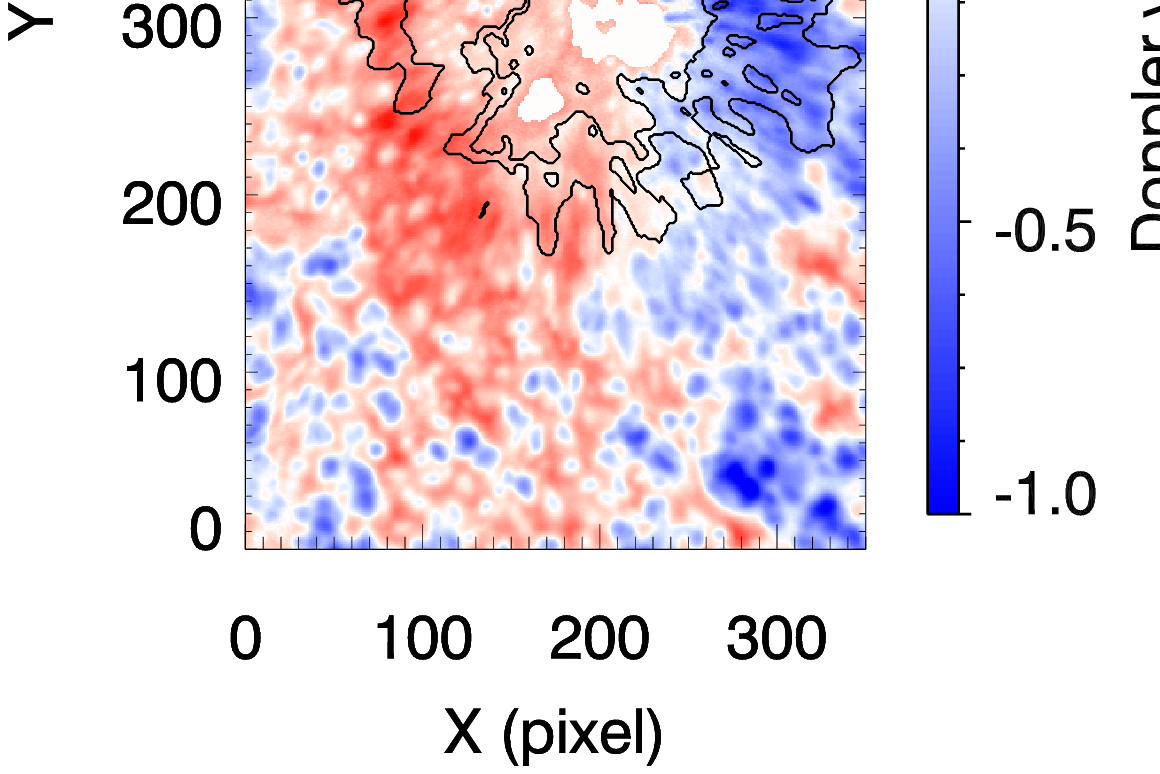}
\includegraphics[trim=110 20 0 75, clip, scale=0.35]{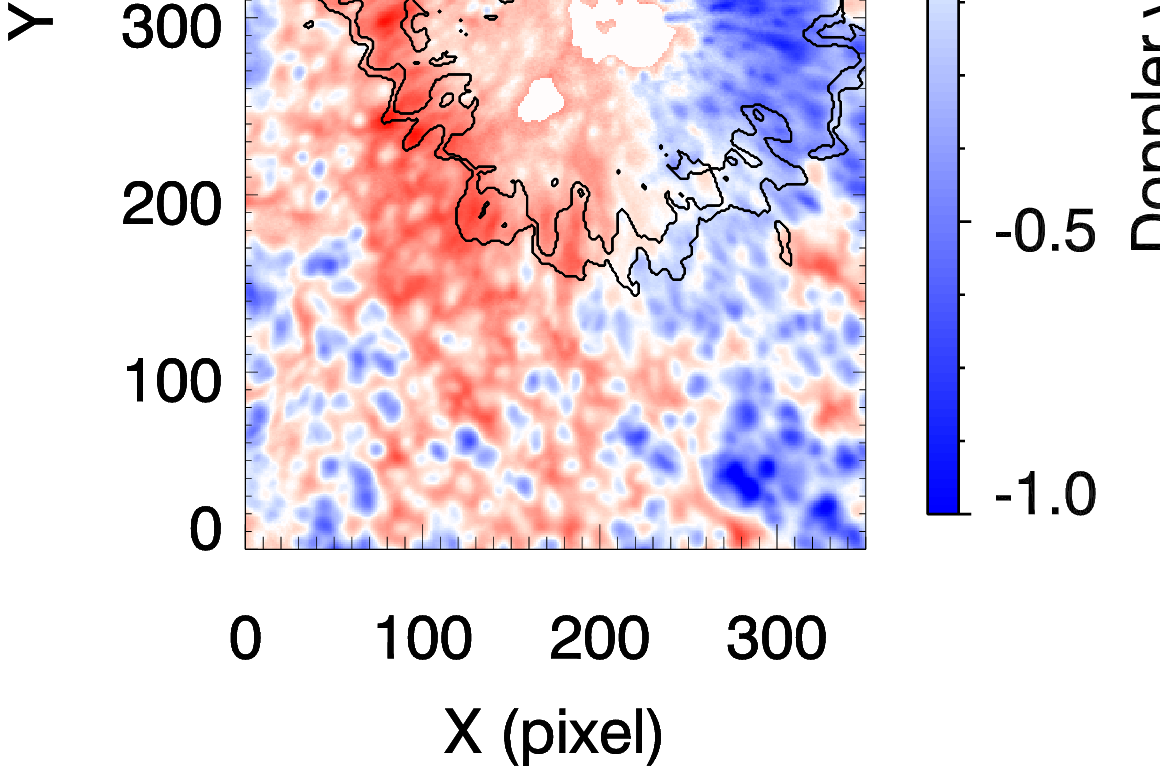}
\caption{Doppler velocity map obtained by the monochromatic images taken on May 18 at 14:42 UT along the Fe I line at 630.25 nm. The positive (red) and negative (blue) values correspond to downflow and upflow velocities. The contours of the features corresponding to the head (left panel), body (central panel) and tail (right panel) of the penumbral filaments are overplotted to the velocity map.}
\label{Fig_new_vel}
\end{center}
\end{figure}

\begin{figure}
\begin{center}
\includegraphics[trim=0 20 160 75, clip, scale=0.35]{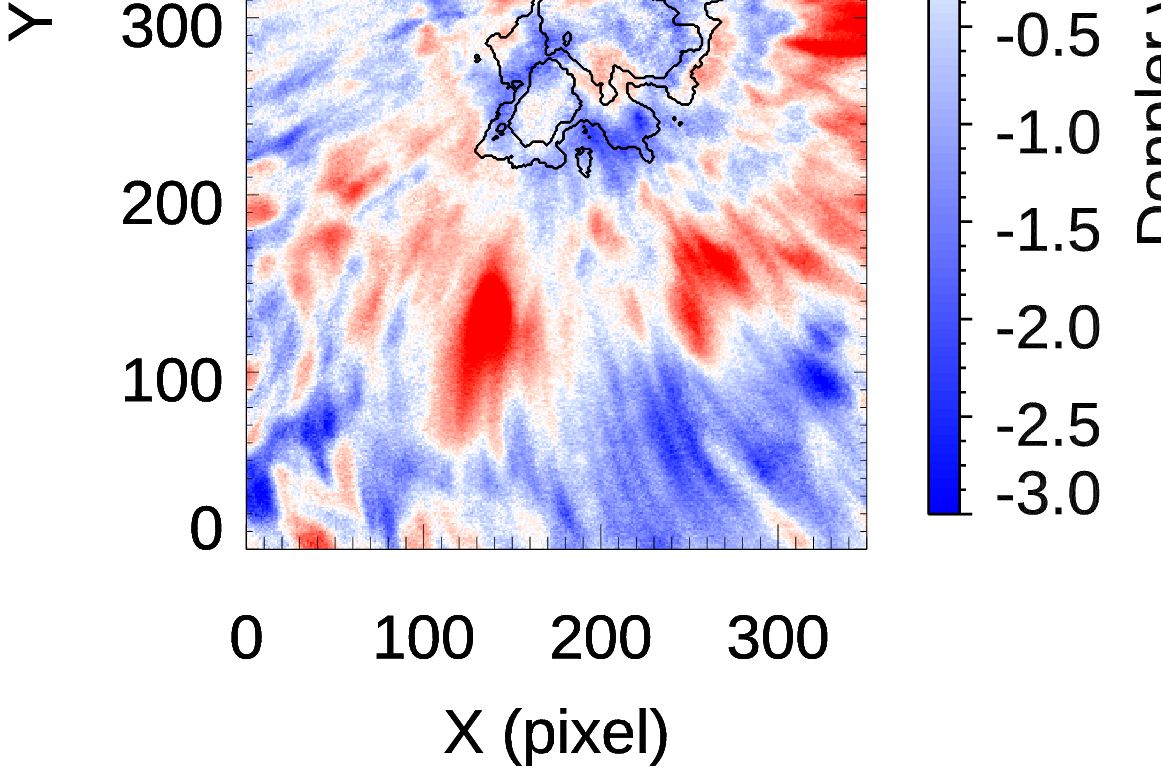}
\includegraphics[trim=110 20 160 75, clip, scale=0.35]{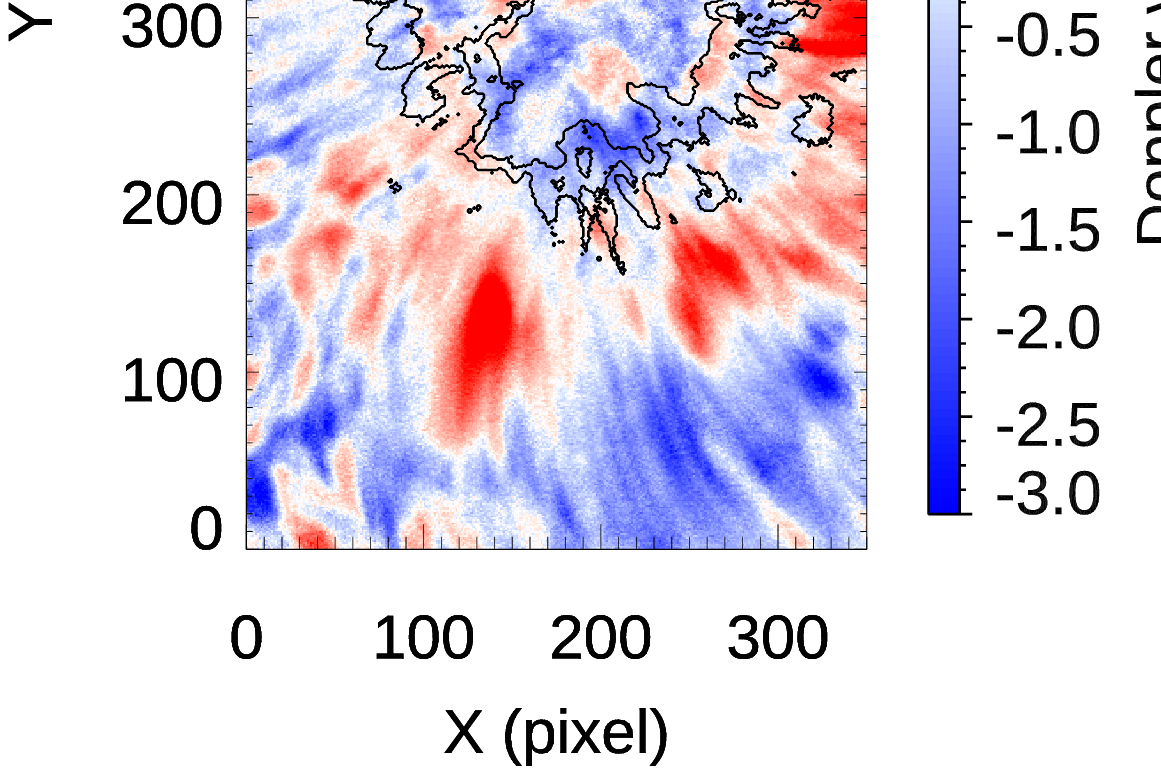}
\includegraphics[trim=110 20 0 75, clip, scale=0.35]{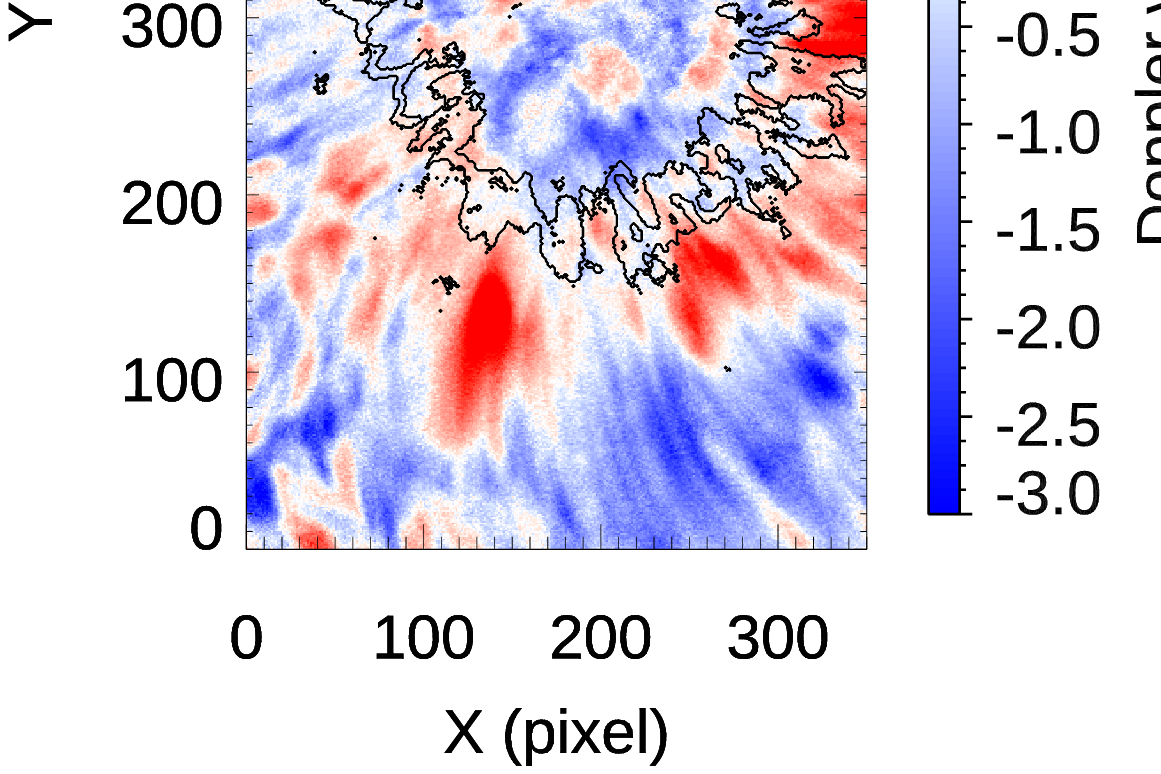}
\caption{Doppler velocity map obtained by the monochromatic images taken on May 18 at 14:42 UT along the H$\alpha$ line. The positive (red) and negative (blue) values correspond to downflow and upflow velocities. The contours of the features corresponding to the head (left panel), body (central panel) and tail (right panel) of the penumbral filaments are overplotted to the velocity map.}
\label{Fig3}
\end{center}
\end{figure}

\begin{figure}
\begin{center}
\includegraphics[trim=0 20 160 75, clip, scale=0.35]{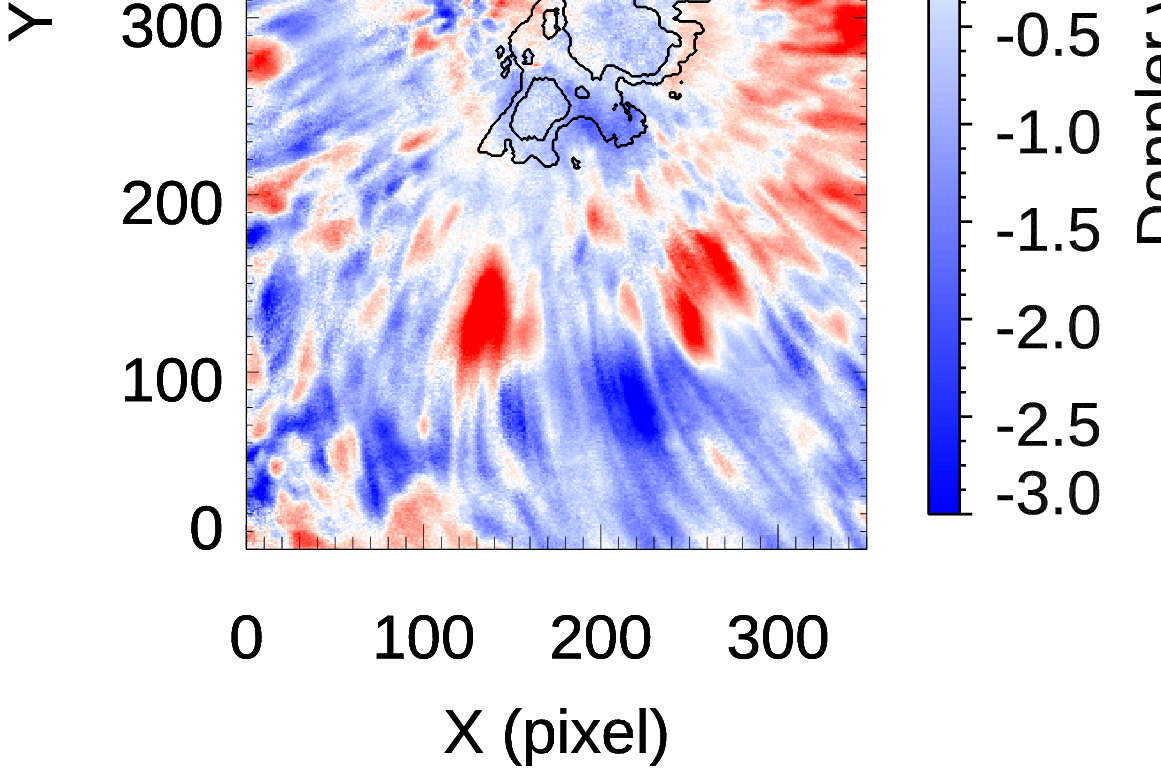}
\includegraphics[trim=110 20 160 75, clip, scale=0.35]{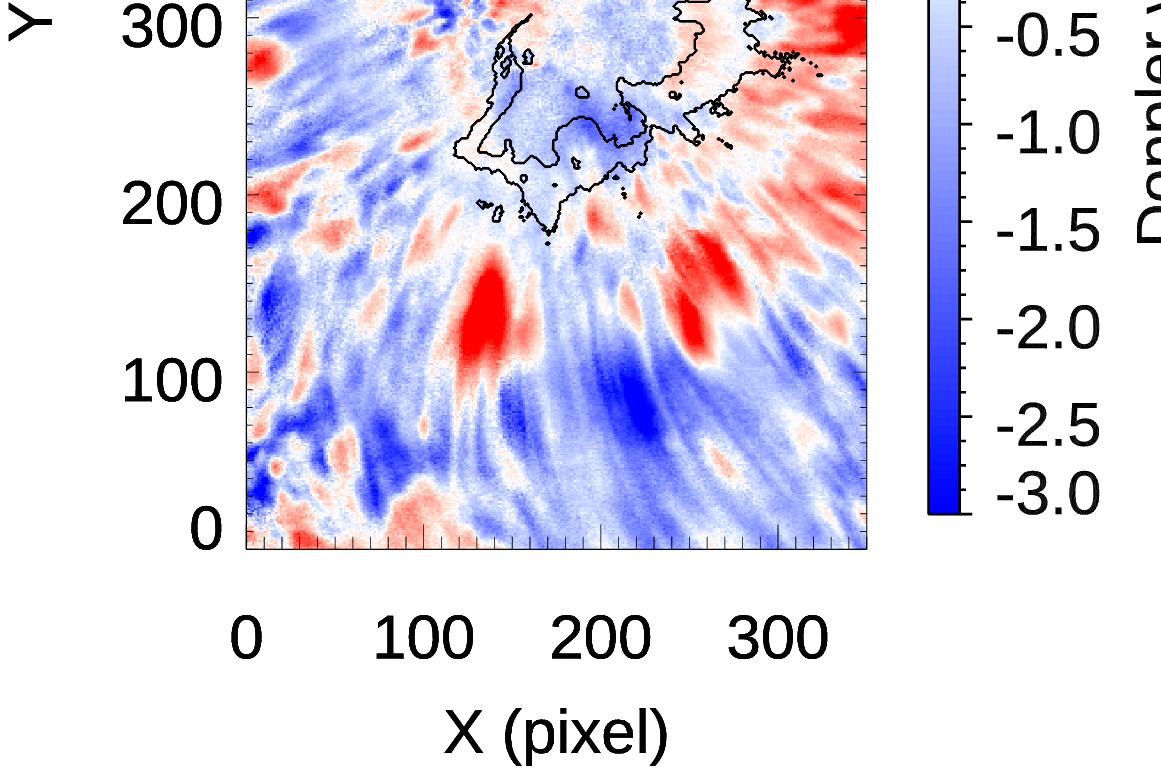}
\includegraphics[trim=110 20 0 75, clip, scale=0.35]{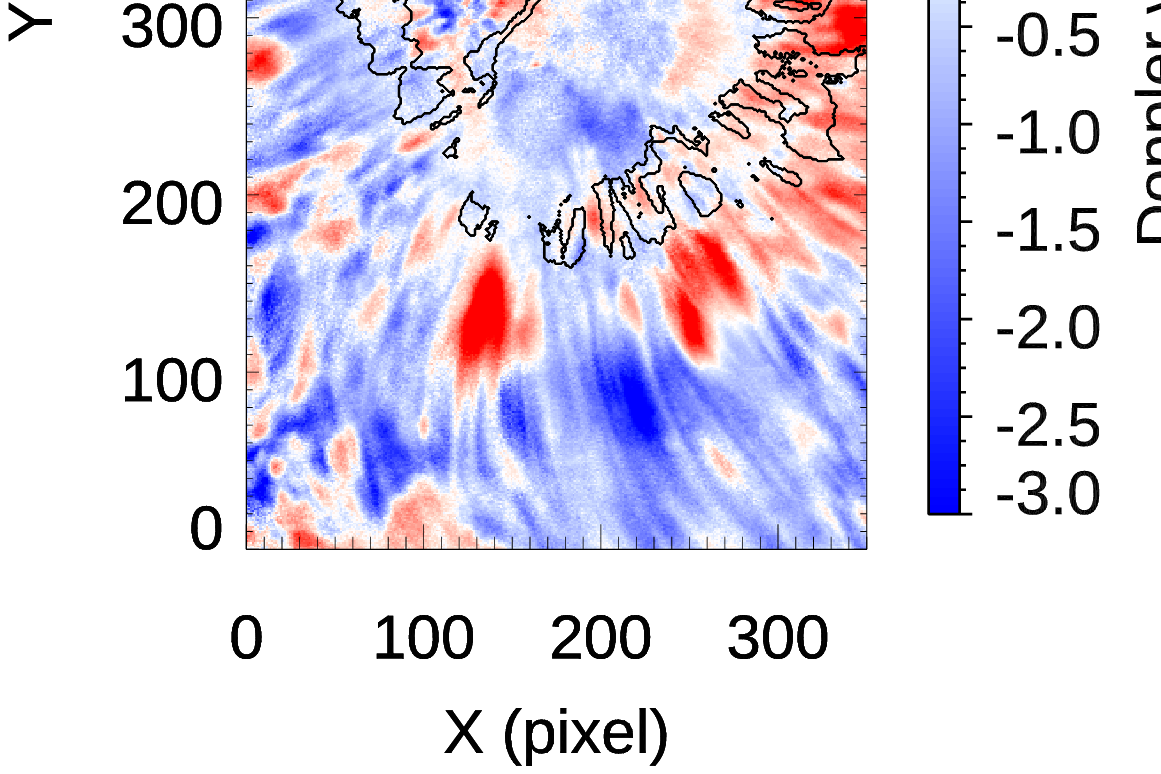}
\caption{Doppler velocity map obtained by the monochromatic images taken on May 18 at 14:42 UT along the Ca II line. The positive (red) and negative (blue) values correspond to downflow and upflow velocities. The contours of the features corresponding to the head (left panel), body (central panel) and tail (right panel) of the penumbral filaments are overplotted to the velocity map.}
\label{Fig4}
\end{center}
\end{figure}

Using the spectroscopic images obtained along the two spectral lines it has been possible to compute some physical parameters which allow us to investigate the behaviour of the plasma along the filaments of the super-penumbra. We reconstructed the profile of each spectral line by fitting in each spatial pixel the signals obtained in the monochromatic images with a linear background and a Gaussian shaped line:
\begin{equation}
I_{j}=A_{0}exp(-\frac{\lambda_{k}-A_{1}}{A_{2}}) + A_{3} + A_{4}\lambda_{k}
\end{equation}
where $I_{j}$ is the observed count rate in pixel $j$ at wavelength $\lambda_{k}$,
$A_{0}$ the height of the Gaussian, $A_{1}$ the centroid wavelength of the Gaussian, $A_{2}$ the $1/e$ half width of the Gaussian, $A_{3}$ the constant term and $A_{4}$ the linear term. The values of velocity along the line of sight in photosphere and chromosphere have been deduced from the Doppler Shift (DS) of the centroid ($A_{1}$) of the line profiles in each spatial point with respect to the center ($\lambda_{0}$) of the corresponding line in the laboratory reference system:
\begin{equation}
v=-\frac{A_{1}-\lambda_{0}}{\lambda_{0}} c
\end{equation}
where $c$ is the light velocity \citep{Sch21}. In particular, we calibrated the phtospheric Dopplergrams assuming that the average LOS velocity value of the sunspot umbra is zero. To identify the pixels belonging to the umbra, we utilized the feature obtained through the SOM technique, which effectively segments the umbra.

Then, the segmentation of the penumbra obtained by the SOM technique provided the opportunity to estimate the values of the DS in different portions of the penumbral filaments by an unsupervised approach.

\section{Results}

\begin{table}
\caption{Doppler velocity estimation using H$\alpha$ and Ca II spectral lines within the features identified through the application of the SOM technique to the same spectral lines. The velocities are expressed in km/s.\label{table1}}
\centering
\begin{tabular}{c c c c c c}
\tableline
\tableline
Feature & Average      & Median     & Maximum    & Maximum & Standard  \\
        &            &            & downflow   & upflow  & deviation \\
\hline
H$\alpha$ & & & & & \\
\hline
Head & -0.57 & -0.59 & +3.29 & -2.70 & 0.73\\
Body & -0.25 & -0.27 & +3-33 & -2.60 & 0.67\\
Tail & -0.01 & +0.05 & +2.65 & -2.28 & 0.63\\
\hline
Ca II & & & & & \\
\hline
Head & -0.20 & -0.14 & +2.46 & -2.00 & 0.61\\
Body & -0.03 & +0.00 & +1.54 & -2.14 & 0.52\\
Tail & +0.32 & +0.14 & +3.68 & -2.95 & 0.88\\
\hline

\tableline
\end{tabular}
\end{table}

\begin{table}
\caption{Doppler velocity estimation using H$\alpha$ and Ca II spectral lines within the features identified through the application of the SOM technique to the Fe I line at 630.25 nm. The velocities are expressed in km/s.\label{table2}}
\centering
\begin{tabular}{c c c c c c}
\tableline
\tableline
Feature & Average      & Median     & Maximum    & Maximum & Standard  \\
        &            &            & downflow   & upflow  & deviation \\
\hline
H$\alpha$ & & & & & \\
\hline
Head & -0.57 & -0.68 & +2.51 & -2.69 & 0.68\\
Body & -0.07 & -0.18 & +5.34 & -2.60 & 0.90\\
Tail & +0.27 & +0.09 & +4.93 & -2.24 & 0.88\\
\hline
Ca II & & & & & \\
\hline
Head & -0.21 & -0.14 & +2.70 & -2.70 & 0.65\\
Body & +0.23 &  0.10 & +6.78 & -2.95 & 0.85\\
Tail & +0.21 & -0.03 & +5.51 & -2.84 & 0.97\\
\hline

\tableline
\end{tabular}
\end{table}

Figure \ref{Fig_new_vel} shows at 603.25 nm the typical Evershed flow, i.e. the shift of the centroid wavelength to the red in the limb-side penumbra and to the blue in the center-side penumbra. Each contour in each panel of Figure \ref{Fig_new_vel} locates the features obtained by the application of the SOM to the same spectral line and corresponding to the head, body, and tail of the penumbral filaments around the sunspot. Looking at the Western side of the sunspot we note that the heads of the penumbral filaments corresponds to velocities closer to zero in comparison to the bodies of the filaments where the negative (blue) velocities are higher in absolute value than 0.5 km/s. While the tail of the penumbral filaments (right panel of Figure \ref{Fig_new_vel}) correspond to regions where several positive (red) patches are visible.

\begin{figure}
\begin{center}
\includegraphics[trim=0 0 100 70, clip, scale=0.55]{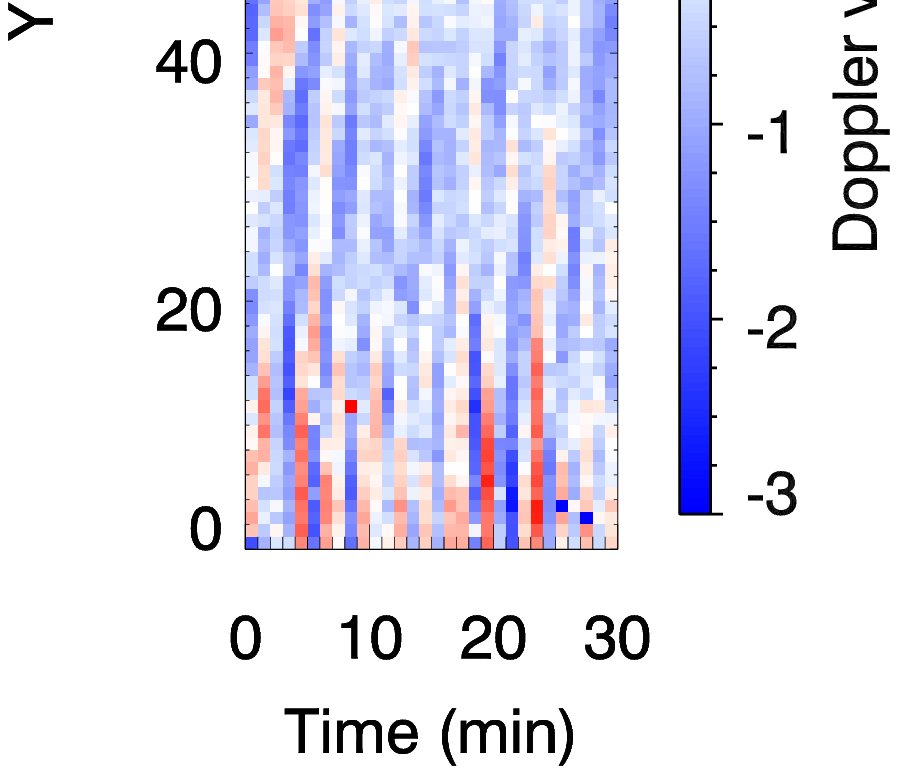}
\caption{This map represents the H$\alpha$ DS variations along the cut delineated by the dashed line in the middle right panel of Figure \ref{Latest}, corresponding to the longitudinal axis of the exemplified filament indicated by the arrow in the top left panel of Figure \ref{Latest}. In the map, the bottom corresponds to the left side, and the top corresponds to the right side of Figure \ref{Latest}. t=0 corresponds to the time when the sequence used in our analysis was acquired.}
\label{time_distance}
\end{center}
\end{figure}

\begin{figure}
\begin{center}
\includegraphics[trim=40 160 160 500, clip, scale=0.55]{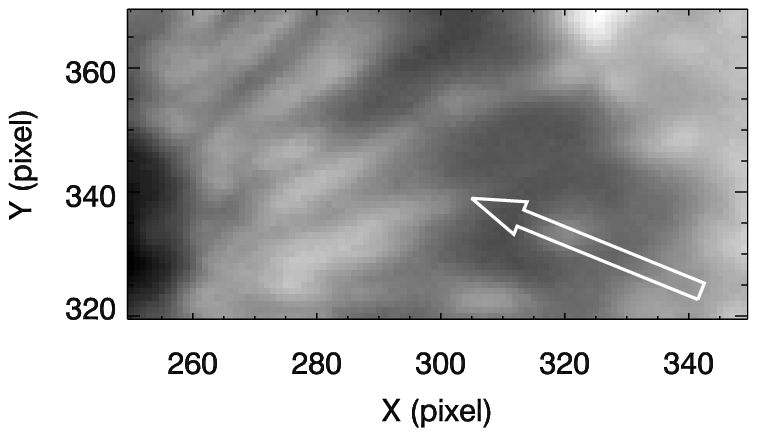}
\includegraphics[trim=115 160 80 500, clip, scale=0.55]{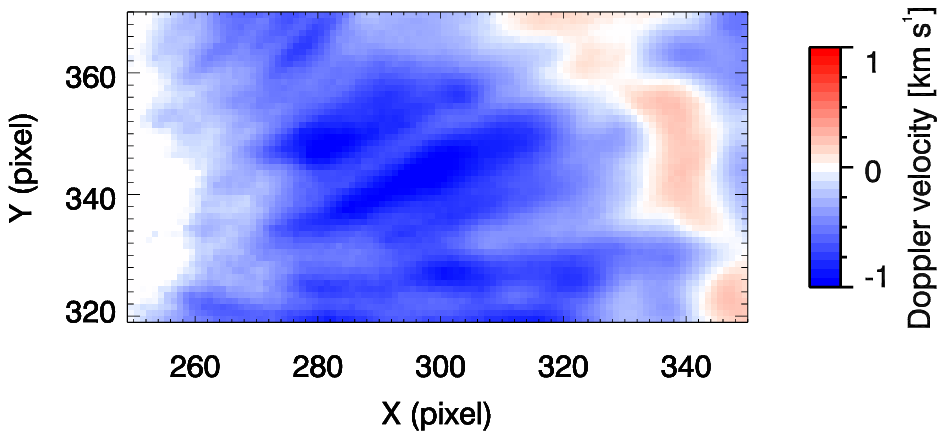}\\
\includegraphics[trim=40 160 160 500, clip, scale=0.55]{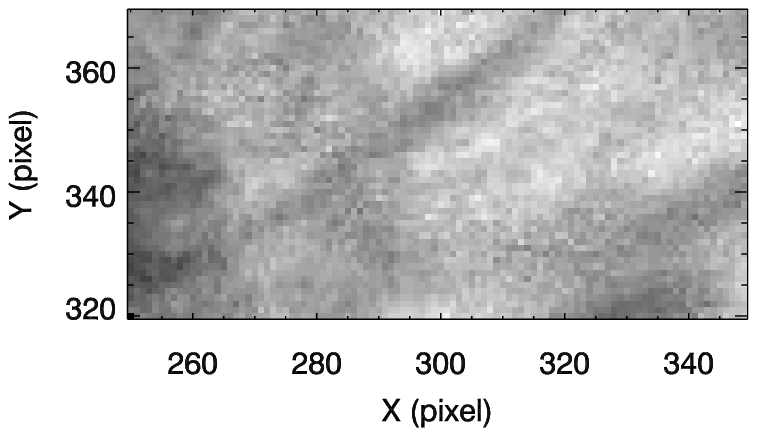}
\includegraphics[trim=115 160 80 500, clip, scale=0.55]{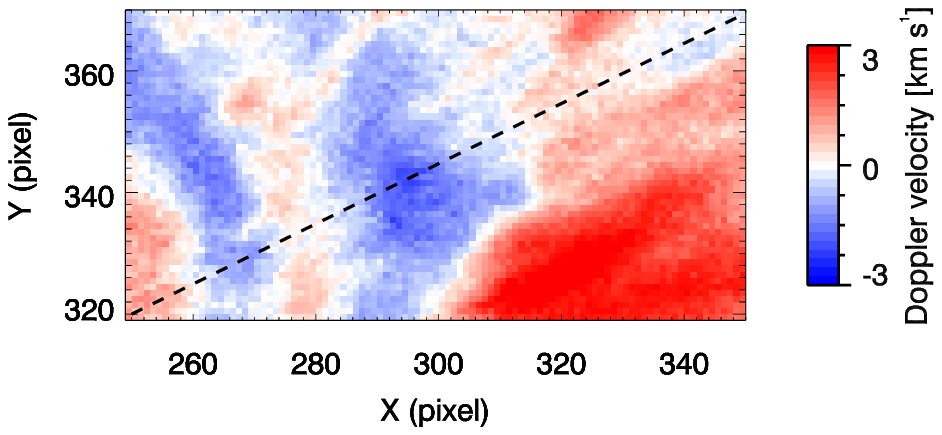}\\
\includegraphics[trim=40 110 160 500, clip, scale=0.55]{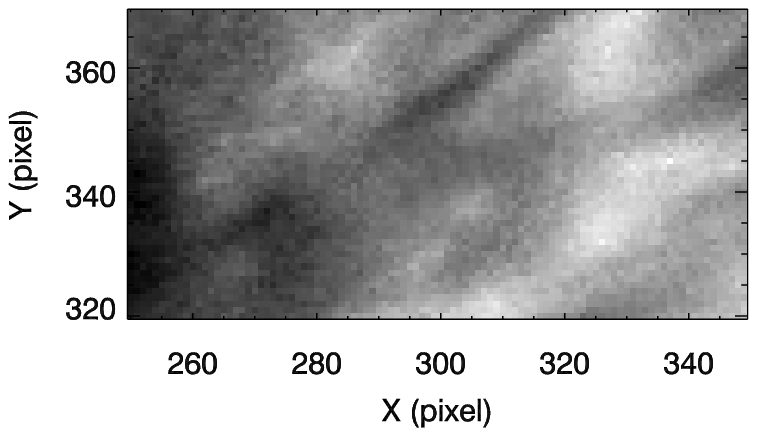}
\includegraphics[trim=115 110 80 500, clip, scale=0.55]{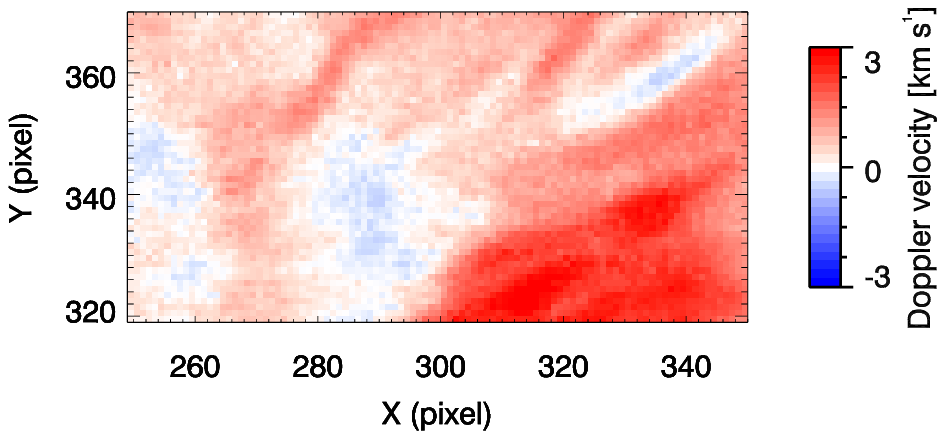}

\caption{Enlargement of the field of view indicated by the white box in the bottom left panel of Figure \ref{Fig0}. In the top left panel, an example filament visible at 630.28 nm is highlighted by the arrow. The left panels display spectral images captured in the core of the Fe I 630.250 nm line (top), H$\alpha$ line (middle), and Ca II line (bottom). The corresponding velocity maps are shown in the right panels.}
\label{Latest}
\end{center}
\end{figure}

DS values for the two chromospheric spectral lines are displayed in Figures \ref{Fig3} (H$\alpha$) and \ref{Fig4} (Ca II), providing valuable insights into the physical processes that govern the penumbra and super-penumbra behavior. {Indeed, these lines allowed investigate the low solar atmosphere spanning from the photosphere by the monochromatic images taken along their wings to the chromosphere by the spectral images taken in the cores of the lines.} Both velocity maps obtained by the two chomospheric lines show a global flow pattern which recalls the super-penumbra fine structure characterized by fibrils with a radial distribution. Looking at the areas where the upflow (blue) or downflow (red) prevail, we note that going from the sunspot center to its edge there is an alternating rings characterized by prevalent opposite directions of the vertical plasma motions.  If we exclude the Eastern side of the sunspot where a decay process is taking place (see \citet{Rom20} for more details), we see clearly the upflows which surround the sunspot near the southern edge of the FOV and in the nothern area located between the sunspot and the facula. Unfortunately, we do not see the portion of this upflows located at the Western side of the sunspot because it is supposedly located out of the FOV. These upflows, together with the downflows located in a ring closer to the center of the spot, can be interpreted as the vertical components of the inverse Evershed flow, which transport the plasma along the super-pemubral fibrils from their outer portions toward the center of the sunspot. We can exclude that these rings characterized by opposite directions of the plasma motions may be ascribed to shock waves produced by photospheric oscillations because we find the same pattern also in the subsequent sequences acquired after the one used in our analysis. In fact, we observe the persistence of this pattern in the time-distance map of H$\alpha$ DS shown in Figure \ref{time_distance}. This map was generated along the cut corresponding to the longitudinal axis of a penumbral filament, as indicated by the dashed line in the middle right panel of Figure \ref{Latest}.

However, this global scenario appears more complex when we analyse more deeply the velocity maps by the application of the SOM technique in the region corresponding to the penumbra. Indeed, the fine DS pattern in the penumbra requires a segmentation to distinguish among the head, body, and tail of the penumbral filaments. For this reason, by applying the masks corresponding to these features onto the Doppler velocity maps, we were able to isolate the three distinct portions of the penumbral filaments. Considering that these features exhibit mixed LOS velocity components, we measured the average and median velocities in these features and we reported them in Table \ref{table1}. We found a decrease in both the average and median upflows in the H$\alpha$ line along the penumbral filaments, from the heads to the tails. We measured an average upflow of -0.57 km/s in the penumbral filament head and an average LOS velocity of about zero at the tails. Instead, using the Ca II line, we found an inversion in the vertical direction of plasma velocity, with the heads of the filaments which exhibit an average velocity of -0.20 km/s and the tails which display an average velocity of +0.32 km/s. The feature corresponding to the body of the filaments was characterized by a null vertical component of plasma velocity. Therefore, both spectral lines reveled velocities consistent with the classical Evershed flow and the more recent MHD simulations. We remark that the standard deviations are of the same order of magnitude of the corresponding average velocities, due to the limits of the data spatial resolution and the DS computation.

To enhance the robustness of our findings, we have also calculated the average DS velocities using the H$\alpha$ and Ca II lines within the features identified through the application of the SOM technique to the Fe I line at 630.25 nm. The obtained values are presented in Table \ref{table2}, and it is evident that they are consistent with the results reported in Table \ref{table1}. Indeed, we found an average upflow in the heads and a downflow in the tails of the penumbral filaments. Some slight differences between the velocities estimated by the application of the segmantations obtained with the chromospheric lines them self or with the Fe I line regard only the average DS of the bodies of the penumbral filaments.

We remark that the photospheric and chromospheric patterns depicted in Figures \ref{Fig_new_vel}, \ref{Fig3} and \ref{Fig4} exhibit significant differences. This divergence arises as a result of the inherent characteristics of the Evershed and inverse Evershed flows, which these respective layers are expected to sample. However, the strong correlation observed in the average velocities, as evident from the comparison in Table \ref{table1} and \ref{table2}, can be attributed to the consistent identification of the heads, bodies, and tails of the penumbra when applying the SOM technique to both the Fe I 630.25 line and the chromospheric lines. This confirms that monochromatic images captured along the wings of the two chromospheric lines suffice for identifying the photospheric component of the penumbra.

From the segmentation of the tails of the penumbral filaments, we also remark that the downflow region surrounding the sunspot appears to be shared by both the classic Evershed flow of the penumbral filaments and the inverse Evershed flow of the super-penumbra fibrils. This is more clearly illustrated in Figure \ref{Latest}, where we focus on a penumbral filament, indicated by the arrow in the top left panel showing the core of the Fe I 630.25 line. Although the photspheric penumbral filament is not discernible in the core of the chromospheric lines (middle and bottom left panels of Figure \ref{Latest}), the velocity maps (right panels of Figure \ref{Latest}) provide a distinct view of the downflow surrounding the sunspot. In fact, the tail of the penumbral filament corresponds to regions where the elongated red patch is visible on the western side of the Fe I 630.25 line velocity map (top right panel of Figure \ref{Latest}). This same area aligns with the downflow region on the western side of the velocity maps for the H$\alpha$ line (middle right panel of Figure \ref{Latest}) and the Ca II line (bottom right panel of Figure \ref{Latest}).

\section{Conclusions}

We applied the SOM technique to study the sunspot penumbra in our spectral dataset, which contains monochromatic images taken along the Ca II 854.2 nm and H$\alpha$ 656.28 nm lines. These observations have provided important insights into the dynamics of sunspots using those chromospheric lines.

\begin{figure}
\begin{center}
\includegraphics[trim=0 0 0 0, clip, scale=0.2]{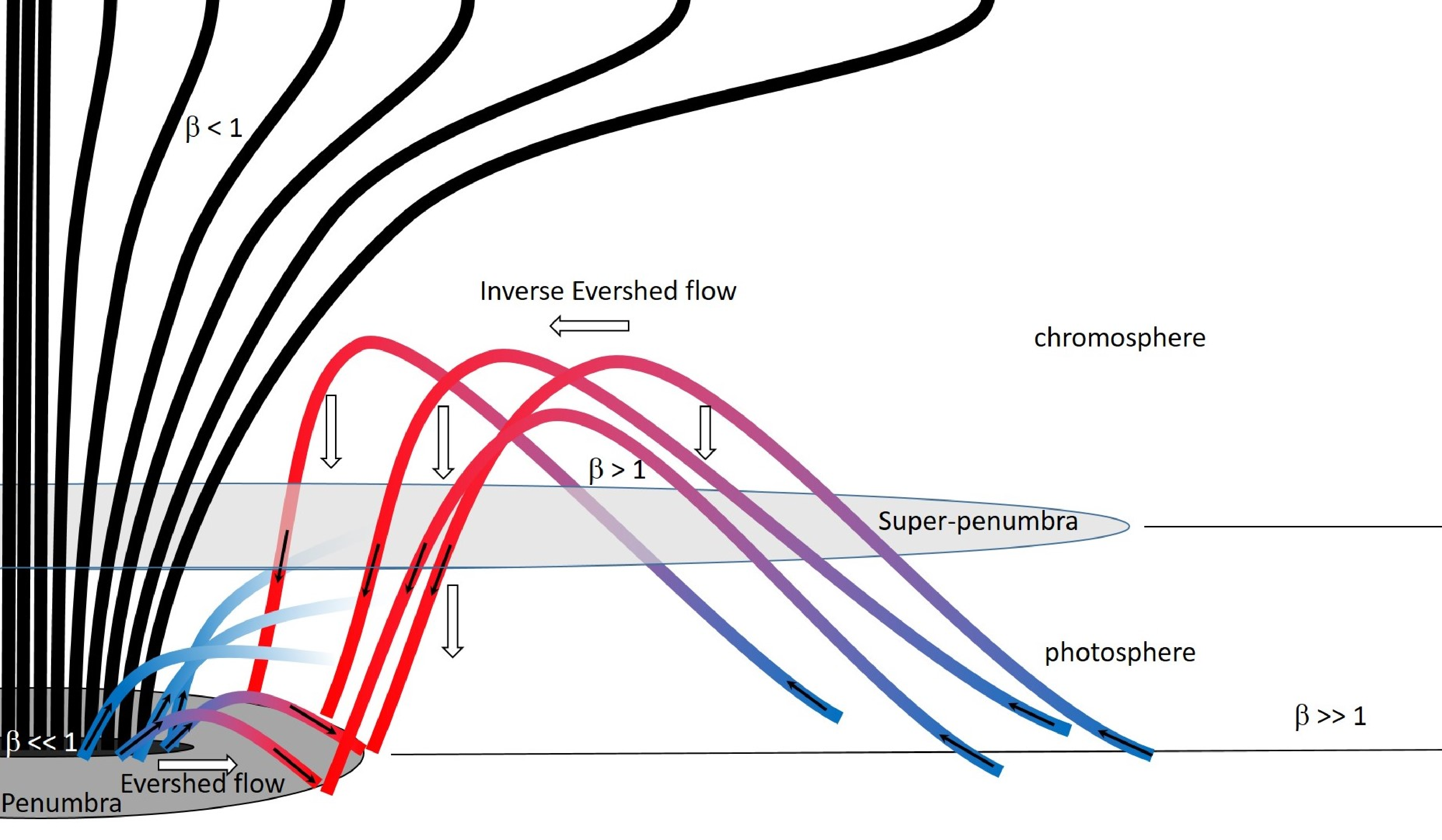}
\caption{Sketch summarizing the obtained results. The black lines represent the umbra magnetic filed lines. Blue and red correspond to plasma upflow and downflow, respectively. The white downward arrows indicate the field lines which may be torn down.}
\label{Fig5}
\end{center}
\end{figure}
 
Our data analysis enabled us to segment the penumbra and measure plasma velocity along the chromospheric regions of the penumbral filaments. We observed the presence of both classic and inverse Evershed flows along the penumbral filaments and super-penumbral fibrils, respectively. Notably, the inverse Evershed flow was observed in the outer region of the sunspot's magnetic field system, but not above the penumbra visible in the photosphere.

The sketch reported in Fig. \ref{Fig5} summarizes the plasma behaviour in the context of the uncombed model of the sunspot penumbra, taking into account our findings. The more horizontal field lines that remain at lower levels of the solar atmosphere are governed by the outwards plasma motion, which is evident through the detection of upflows in the heads and downflows in the tails of the  penumbral filaments observed by the spectral images taken in the continuum near the Ca II line. Other field lines with a higher inclination are still capable of exhibiting outward motion through the upflow observed in the inner region of the penumbra. However, in this case the upflow values decrease along the bodies of the penumbral filaments, as observed by the analysis of the monochromatic images taken along the H$\alpha$ line. Instead, the inverse Evershed flow seem to involve other field lines that connect the tails of the penumbra filaments to the outer and higher portions of the sunspot magnetic field system. Taking into account, that we found the two opposite Evershed regimes working next to each other, without overlapping, this scenario explains the alternation of up-down-up motions observed from the umbra-penumbra edge to the outer part of the super-penumbra. In particular, we can ascribe the downflow observed in the outer edges of the sunspot to both the palsma accelerated downwards at the tails of the penumbral filaments by the classic Evershed flow and the material flowing down through the upper layers of the solar atmosphere, corresponding to the inner portions of the inverse Evershed flow.

However, it is important to note that we cannot exclude the possibility that the inverse Evershed flow is attributed to the magnetic field dragging the plasma downward. According to the model derived by \cite{Bou17}, which aims to understand the plasma dynamics of the penumbra based on the plasma beta stratification in the solar atmosphere, we are aware that the decrease in beta (the ratio of thermal pressure to magnetic pressure) with height above the super-penumbra permits the downward flow of plasma along the magnetic field lines. This occurs in a regime where beta $>$ 1, resembling the behavior of coronal material that cools due to radiative losses. Indeed, as the magnetic topology of the magnetic field above a sunspot looks like a funnel-shaped flux tube that expands with height in the chromosphere and transition region, the field becomes more inclined above the surrounding of a sunspot than above its center. This implies that the variability of beta is higher in the upper layers around the sunspot umbra. Then, we can also consider that the inclined field in the same beta regime is gravitationally less stable than the more vertical one. Therefore, the inclined field lines may be torn down by continuous mass-loading from above (see the white downward arrows in Fig. \ref{Fig5}).  This process can start earlier at the boundaries of a sunspot, during the penumbra formation, not only because of the less vertical field inclination but also because of the lower total flux density at the surrounding of a sunspot (where beta reaches unity higher above the photosphere than directly above the sunspot where the field is stronger and rather vertical). Thus, once a penumbra forms by the magnetic field being dragged down to the photosphere, neighboring field lines might also experience an additional force dragging them down. 

The plasma downflow observed in the super-penumbra could also be ascribed to magnetic reconnection process between the falling field lines and preexisting magnetic fields in the lower atmosphere outside of the sunspot. Due to the fact that these field lines are initially connected to lower-pressure atmospheric layers, they need to be filled with additional mass when reconnecting from a lower thermal pressure regime (in the upper atmosphere) to photospheric layers outside of the sunspot. The equilibration of this pressure imbalance requires a short-lived plasma flow along the penumbral field toward the sunspot. This happens especially during the early penumbra formation phase. As a result, a penumbra may form through field lines that are filled by cooling material down flowing from the corona \citep{Rom13, Rom14, Rom17}. Actually, we could conclude that the inverse Evershed flow may be interpreted as a continuation of the so-called counter Evershed flow, i.e., the DS pattern observed near a pore when the penumbra formation is taking place \citep{Mur16}. In this case, one would expect that additional penumbra forms around existing penumbra, which is what is often observed. Therefore, the signature of growing penumbra confirms that the field inclination continues to become more horizontal, and plasma simultaneously falls from above during the penumbra life.

For these reasons, we think that our results are important in the context of sunspot formation and evolution. However, they suggest that a better understanding of the physical properties of sunspot penumbrae is needed, particularly in the chromospheric layer where super-penumbrae are observed. Further investigation is needed to fully comprehend the behavior of plasma in the peculiar configuration of the magnetic field of the sunspot penumbrae. Furthermore, the application of the SOM technique to high-resolution images has proven to be a valuable tool for segmenting and analyzing solar images in a precise and unsupervised manner. However, we believe that its application to solar full-disc images with lower spatial resolution may have significant implications for the segmentation of features, especially regarding their utility for space weather purposes.

\acknowledgments

The authors wish to thank the DST staff for its support during the observing campaigns. The research leading to these results has received funding from the European Union’s Horizon 2020 research and innovation programme under grant agreement no. 824135 (SOLARNET project). This work was supported by the Italian MIUR-PRIN 2017 on Space Weather: impact on circumterrestrial environment of solar activity and by the Space Weather Italian COmmunity (SWICO) Research Program.
A special thanks to MOSAICo projects (Metodologie Open Source per l'Automazione Industriale e delle procedure di CalcOlo in astofisica) funded by Italian MISE, Ministero Sviluppo Economico).   



{\it Facilities:} \facility{SDO (AIA, HMI), SOHO (LASCO), STEREO, GONG , RHESSI....}.





\clearpage



\end{document}